\documentclass[prx,twocolumn,floatfix,preprintnumbers,amsmath,amssymb,
superscriptaddress, a4paper]{revtex4-2}
\usepackage{geometry}
\usepackage{graphicx}% Include figure files
\usepackage{dcolumn}% Align table columns on decimal point
\usepackage{bm}% bold math
\usepackage{color}
\usepackage{xcolor} % Add nice colors to comment....
\usepackage{xspace}
\usepackage{amsmath}
\usepackage{appendix}
\usepackage{multirow} 
\usepackage{subfigure}

\usepackage{textcomp} % Use with pdfLaTeX

\graphicspath{{./}{figure/}}
\usepackage[colorlinks,plainpages=false,linkcolor=blue,urlcolor=blue,citecolor=blue,pdfpagemode=UseNone,pdfstartview=FitBH]{hyperref}
\usepackage{dcolumn,graphicx,color,booktabs,microtype,afterpage}
\newcommand{\chemCPA}{(C$_5$H$_9$NH$_3$)$_2$CuBr$_4$}

\begin{document}

%\title{Static and Dynamic Lattice Effects on Quantum Magnetism in a Metal-Organic Framework \chemCPA}
\title{Magnetic and phononic dynamics in the two-ladder quantum magnet \chemCPA}

\author{J.~Philippe}
\email{jonas.philippe@psi.ch}
\affiliation{PSI Center for Neutron and Muon Sciences, CH-5232 Villigen-PSI, Switzerland}
\affiliation{Physik-Institut, Universit\"{a}t Z\"{u}rich, Winterthurerstrasse 190, CH-8057 Z\"{u}rich, Switzerland}

\author{F.~Elson}
\affiliation{Department of Applied Physics, KTH Royal Institute of Technology, SE-106 91 Stockholm, Sweden}

\author{T.~Arh}
\affiliation{PSI Center for Neutron and Muon Sciences, CH-5232 Villigen-PSI, Switzerland}

\author{S.~Sanz} 
\affiliation{Peter Grünberg Institute, Electronic Properties (PGI-6), Forschungszentrum Jülich, 52425 Jülich, Germany}

\author{M.~Metzelaars} 
\affiliation{Peter Grünberg Institute, Electronic Properties (PGI-6), Forschungszentrum Jülich, 52425 Jülich, Germany}
\affiliation{Institute of Inorganic Chemistry, RWTH Aachen University, 52056 Aachen, Germany}

\author{D.~W.~Tam}
\affiliation{Department of Applied Physics, KTH Royal Institute of Technology, SE-106 91 Stockholm, Sweden}

\author{O.~K.~Forslund}
\affiliation{Physik-Institut, Universit\"{a}t Z\"{u}rich, Winterthurerstrasse 190, CH-8057 Z\"{u}rich, Switzerland}
\affiliation{Department of Physics and Astronomy, Uppsala University, Box 516, SE-751 20 Uppsala, Sweden}

\author{O.~Shliakhtun} 
\affiliation{PSI Center for Neutron and Muon Sciences, CH-5232 Villigen-PSI, Switzerland}
\affiliation{Physik-Institut, Universit\"{a}t Z\"{u}rich, Winterthurerstrasse 190, CH-8057 Z\"{u}rich, Switzerland}

\author{C.~Jiang}
\affiliation{Department of Applied Physics, KTH Royal Institute of Technology, SE-106 91 Stockholm, Sweden}

\author{J.~Lass}
\affiliation{PSI Center for Neutron and Muon Sciences, CH-5232 Villigen-PSI, Switzerland}

\author{M.~D.~Le}
\affiliation{ISIS facility, Rutherford Appleton Laboratory, Chilton, Didcot OX11 0QX Oxfordshire, UK}

\author{J.~Ollivier}
\affiliation{Institut Laue Langevin, BP156, 38042 Grenoble, France}

\author{P. Bouillot} 
\affiliation{Department of Quantum Matter Physics, University of Geneva, Quai Ernest-Ansermet 24, CH-1211 Geneva, Switzerland}

\author{T. Giamarchi}
\affiliation{Department of Quantum Matter Physics, University of Geneva, Quai Ernest-Ansermet 24, CH-1211 Geneva, Switzerland}

\author{M.~Bartkowiak}
\affiliation{PSI Center for Neutron and Muon Sciences, CH-5232 Villigen-PSI, Switzerland}

\author{D.~G.~Mazzone}
\affiliation{PSI Center for Neutron and Muon Sciences, CH-5232 Villigen-PSI, Switzerland}

\author{P.~K\"{o}gerler}
\affiliation{Institute of Inorganic Chemistry, RWTH Aachen University, 52056 Aachen, Germany}

\author{M.~M\aa nsson}
\affiliation{Department of Applied Physics, KTH Royal Institute of Technology, SE-106 91 Stockholm, Sweden}

\author{A.~M.~Läuchli}
\affiliation{PSI Center for Scientific Computing, Theory and Data, CH-5232 Villigen-PSI, Switzerland}
\affiliation{Institute of Physics, Ecole Polytechnique F\'ed\'erale de Lausanne (EPFL), CH-1015 Lausanne, Switzerland}

\author{Y.~Sassa}
\affiliation{Department of Applied Physics, KTH Royal Institute of Technology, SE-106 91 Stockholm, Sweden}
\affiliation{Department of Physics, Chalmers University of Technology, SE-412 96 G\"{o}teborg, Sweden}

\author{M.~Janoschek}
\affiliation{PSI Center for Neutron and Muon Sciences, CH-5232 Villigen-PSI, Switzerland}
\affiliation{Physik-Institut, Universit\"{a}t Z\"{u}rich, Winterthurerstrasse 190, CH-8057 Z\"{u}rich, Switzerland}

\author{B.~Normand}
\affiliation{PSI Center for Scientific Computing, Theory and Data, CH-5232 Villigen-PSI, Switzerland}

\author{G.~Simutis}
\email{gediminas.simutis@psi.ch}
\affiliation{PSI Center for Neutron and Muon Sciences, CH-5232 Villigen-PSI, Switzerland}
\affiliation{Department of Physics, Chalmers University of Technology, SE-412 96 G\"{o}teborg, Sweden}

\date{\today}

\begin{abstract}
In quantum magnetic materials it is common to observe both static and dynamic lattice effects on the magnetic excitation spectrum. Less common is to find that the magnetic correlations have a signficant impact on the phonon spectrum. Can such an interplay occur in a structurally soft system with comparable elastic and magnetic energy scales? Here we study the metal-organic material \chemCPA~(Cu-CPA), in which an explanation of the low-lying excitations depends crucially on a full understanding of both the spin and lattice subsystems. We report high-resolution neutron spectroscopy enabled by large, deuterated single-crystals that reveal how both sectors are affected by the recently discovered structural phase transition. By measuring over several Brillouin zones, we disentangle the vibrational contribution to the spectrum in order to obtain an accurate estimate of the quasi-one-dimensional magnetic signal. The low-energy magnetic excitations are dominated by two gaps, $\Delta_b = 0.41$~meV and $\Delta_a = 0.55$~meV, which contribute with equal intensity ratios, confirming that Cu-CPA realizes a two-ladder spin Hamiltonian, and we deduce the magnetic interaction parameters of both ladders. The phonon spectrum contains a highly localized mode at an anomalously low-energy around 2 meV. This characteristic frequency drops by approximately 5\% as magnetic correlations become established with decreasing temperature, and we connect this behavior with the location and structure of the cyclopentylammonium rings. 
\end{abstract}

\maketitle{}

\section{Introduction}
\label{sintro}

Insulating materials formed from magnetic ions with spin $S = 1/2$ create an excellent platform for the experimental study of many-body quantum physics. Because these materials have only a small number of well defined and short-ranged interactions, they can realize paradigm low-dimensional spin models whose analysis was once the province only of mathematical physics. This is particularly true in one-dimensional (1D) systems, whose gapless phases can realize integrable models and quantum field theories \cite{Mikeska2004}. In addition to nearest-neighbor spin chains of Heisenberg, Ising, or XY type, 1D models with more than one parameter include alternating chains and chains with next- or further-neighbor interactions, as well as quantum spin ladders \cite{dagotto_surprises_1996,giamarchi_coupled-ladders_1999}. Two-leg ladders form a class of quantum systems whose field-induced ground states and excitation spectra are controlled by a single parameter, the interaction ratio $\alpha = J_\mathrm{leg}/{J_\mathrm{rung}}$ \cite{Schmidiger2013}.

%% FIG 1
\begin{figure*}[t]
    \centering
    \includegraphics[width = \textwidth]{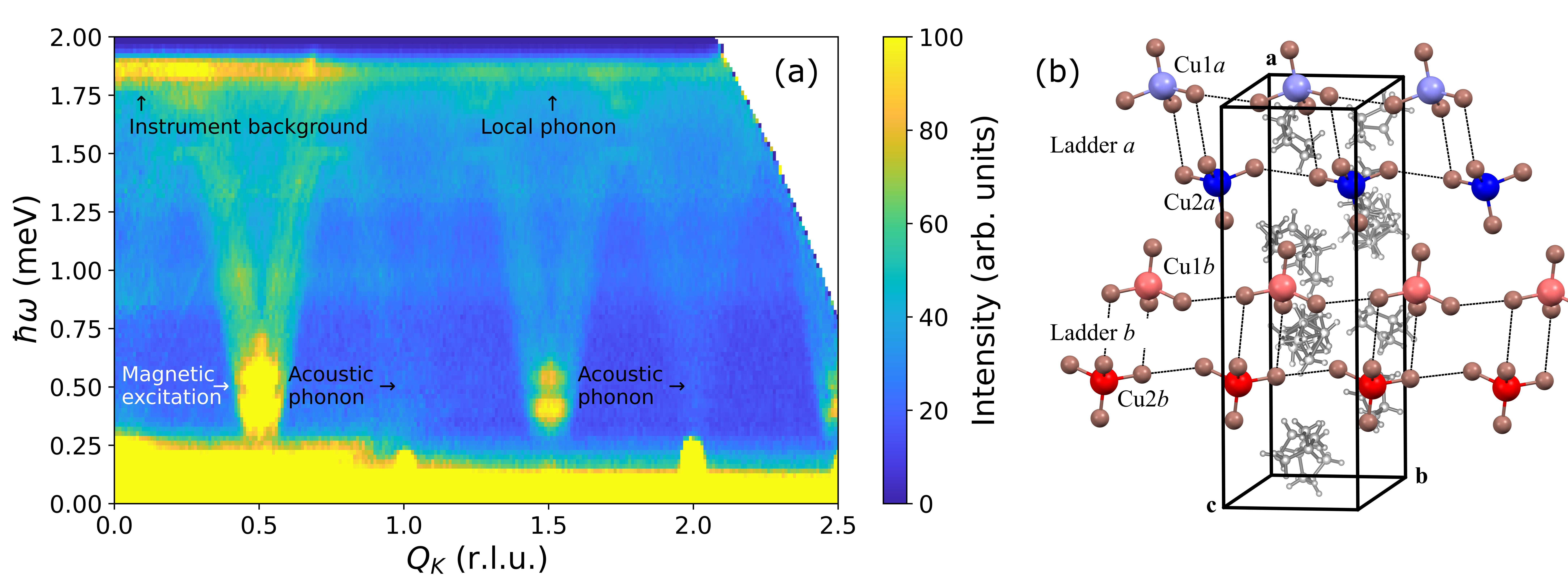}
    \caption{ \label{fig:excitation_spectrum_5p5AA}{\bf Excitation spectrum and lattice structure of Cu-CPA.} (a) Full excitation spectrum measured at $T = 0.05$~K at the IN5 instrument, shown with no background subtraction. The measurements were performed with $E_i = 2.70$~meV and the data were integrated perpendicular to the ladder direction over the ranges $Q_H \in [-0.85, 0.85]$~r.l.u.~and $Q_L \in [-2, 4]$~r.l.u. The dispersive modes with high intensity at half-integer values of $Q_K$ are the magnetic (``triplon'') excitations of the $S = 1/2$ antiferromagnetic Heisenberg spin ladders. The dispersive modes appearing at integer values of $Q_K$ are acoustic phonons, which have low intensity at this measurement temperature. Intensity contributions arising from a low-lying local (optical) phonon and from the instrument background are also marked.
    (b) Low-temperature structure of Cu-CPA, showing the two inequivalent spin ladders (denoted as ``Ladder $a$'' and ``Ladder $b$''). For clarity, only half of the crystallographic c-direction is shown. Following the nomenclature introduced in the structural study of this material \cite{philippe_metal-organic_2024}, the lighter and darker red and blue colors denote inequivalent Cu$^{2+}$ ions on each ladder rung. Here we have shown the cyclopentylammonium (CPA) molecules in gray, projected behind the copper-tetrabromine units, in order to enhance the visibility of the ladders.}
\end{figure*}

Early spin-ladder compounds were based on cuprates \cite{hiroi1991, azuma1994, ishida1994, kojima1995, hiroi1995, mayaffre1998, eccleston1998} and hence had the two properties that $\alpha \approx 1$ and the interaction parameters were well over 100 meV, requiring larger than available laboratory magnetic fields to tune them. Metal-organic materials with a wide variety of ligands provided both geometrical flexibility and, because of the large organic spacer groups, the possibility of very small interaction parameters (meaning in the 1 meV range or truly vanishing \cite{willett_structure_2004}). Several early metal-organic spin ladders, most notably (C$_5$H$_{12}$N)$_2$CuBr$_4$~(BPCB) and (C$_5$H$_{12}$N)$_2$CuCl$_4$~(BPCC) \cite{ruegg2008,Ward2013}, realized the strong-rung regime (small $\alpha$) and therefore could be described very well by bond-operator methods, both in their gapped phases \cite{ward_bound_2017} and in their field-induced gapless ones \cite{Klanjsek2008,thielemann2009a,wehinger2025}. Strong-leg ladders have turned out to be less common, and to date only the compound (C$_7$H$_{10}$N$_2$)$_2$CuBr$_4$~(DIMPY) has been studied in detail in experiment \cite{Schmidiger2011, Schmidiger2012, Schmidiger2013M}, while numerical methods based on exact diagnalzation (ED) and matrix-product states (MPS) have allowed the unbiased numerical calculation of complete excitation spectra at all $\alpha$ and all applied magnetic fields \cite{Schmidiger2013,bouillot2011}. 

To date the treatment of the phonon contribution to the excitation spectrum of metal-organic magnets has been treated in the same way as inorganic materials, meaning that no optical phonons are expected at 1 meV energies and the thermodynamic contributions from phononic excitations have been modelled using the Debye form. However, important properties of phonons in metal-organic frameworks are first that their energies can be rather low, due to the presence of large organic groups, and second that these groups can host many near-degenerate and low-lying phonon modes. Although certain anomalies persist in the body of experimental data on materials such as DIMPY \cite{Jeong2017},the possibility of a connection with such phonons has yet to be considered.

Here we investigate a metal-organic spin-ladder compound, Bis(cyclopentylammonium)tetrabromocuprate [\chemCPA\ or (CPA)$_2$CuBr$_4$, which we abbreviate Cu-CPA] \cite{willett_structure_2004}, in which a detailed consideration of the phonon contribution is integral to the complete understanding of the low-energy spectrum. Cu-CPA was identified initially as a strong-leg spin ladder \cite{willett_structure_2004}, based on magnetic susceptibility and X-ray diffraction measurements performed on powder samples. However, it was discovered recently that Cu-CPA undergoes a structural phase transition at 113 K (119 K in the deuterated analog), which splits the magnetic contribution into equally weighted components arising from two inequivalent spin ladders \cite{philippe_metal-organic_2024}. This transition ensures that certain optical phonons in Cu-CPA lie particularly low in energy, and hence make strong contributions to the spectral and thermodynamic properties. 

The structure of this article is as follows. In Sec.~\ref{smm} we introduce the structural properties of \chemCPA~and their consequences for thermodynamic measurements made to date. In Sec.~\ref{sres} we describe our inelastic neutron scattering (INS) measurements, the analysis by which we achieve an unambiguous separation of the magnetic and lattice contributions, and our determination of the magnetic interaction parameters of both ladders. In Sec.~\ref{ssh} we reanalyze the specific-heat data of Cu-CPA to identify the full phonon contribution. In Sec.~\ref{sdis} we discuss the physical origin of the anomalous low-energy phononic properties and the mechanism for the clear effects of strong magnetic correlations on the phonon spectrum, concluding with a brief summary. 

\section{Material and Methods}
\label{smm}

To obtain high-quality INS spectra, large single crystals of Cu-CPA were grown from solution. Starting with the synthesis method used to discover the material \cite{willett_structure_2004}, we have optimized the growth process \cite{philippe_metal-organic_2024} to obtain crystals that extend up to linear sizes of 25~mm. These grow as platelets, with the other two dimensions spanning 2--4.5~mm and 0.5--2~mm; a typical crystal is shown in App.~\ref{app:sample}. Microscopically, at low temperature (85 K) the compound realizes a monoclinic P112$_1$ space group with a = 23.965 \AA, b = 8.0765 \AA, c = 18.282 \AA, and $\gamma$ = 90.35$^\circ$. Because the material has a rather low density (2~g\,cm${}^{-3}$) \cite{willett_structure_2004} and one Cu$^{2+}$ ion provides the minimal spin of $S = 1/2$, it was necessary to coalign multiple crystals for our spectroscopic studies; two of these crystal assemblies are shown in App.~\ref{app:sample}.

INS measurements were performed at a number of neutron sources. On the Continuous Angle Multiple Energy Analysis (CAMEA) multiplexing spectrometer at the PSI (Villigen, Switzerland) \cite{lass_commissioning_CAMEA_2023} we used an assembly of twelve deuterated single crystals with a total mass of 731.5(1)~mg, coaligned to within 2.5$^\circ$, and measured at two incident energies of 3.6 and 5.4~meV. These experiments allowed us to observe magnetic excitations and hence provided characterization information essential for our later studies, but the very large lattice constant of Cu-CPA (leading to the presence of many Bragg peaks on the detectors) caused significant spurious background scattering (``Currat-Axe spurions'') arising when the INS angle happens to coincide with a Bragg reﬂection of the analyzer or the monochromator.

Time-of-flight experiments do not use crystals to analyze the energy transfer, and for this reason our subsequent high-resolution study was performed on the IN5 spectrometer \cite{ollivier_IN5_2011} at the ILL (Grenoble, France), using a second batch of five large, deuterated single crystals with a total mass of 968.7(1)~mg and co-aligned within 1.0$^\circ$. Because the two-leg ladders are aligned along the crystallographic $b$ axis [Fig.~\ref{fig:excitation_spectrum_5p5AA}(b)], both of these sample assemblies were oriented such that the $b$ and $c$ directions were in the scattering plane (and the $a$ direction out of it). Additionally, we prepared a different sample consisting of seven deuterated single crystals with a total mass of 403.9(1)~mg, coaligned to within 1.5$^\circ$, and oriented with $a$ and $b$ in the scattering plane. 

\begin{figure*}[t]
    \includegraphics[width={\textwidth}]{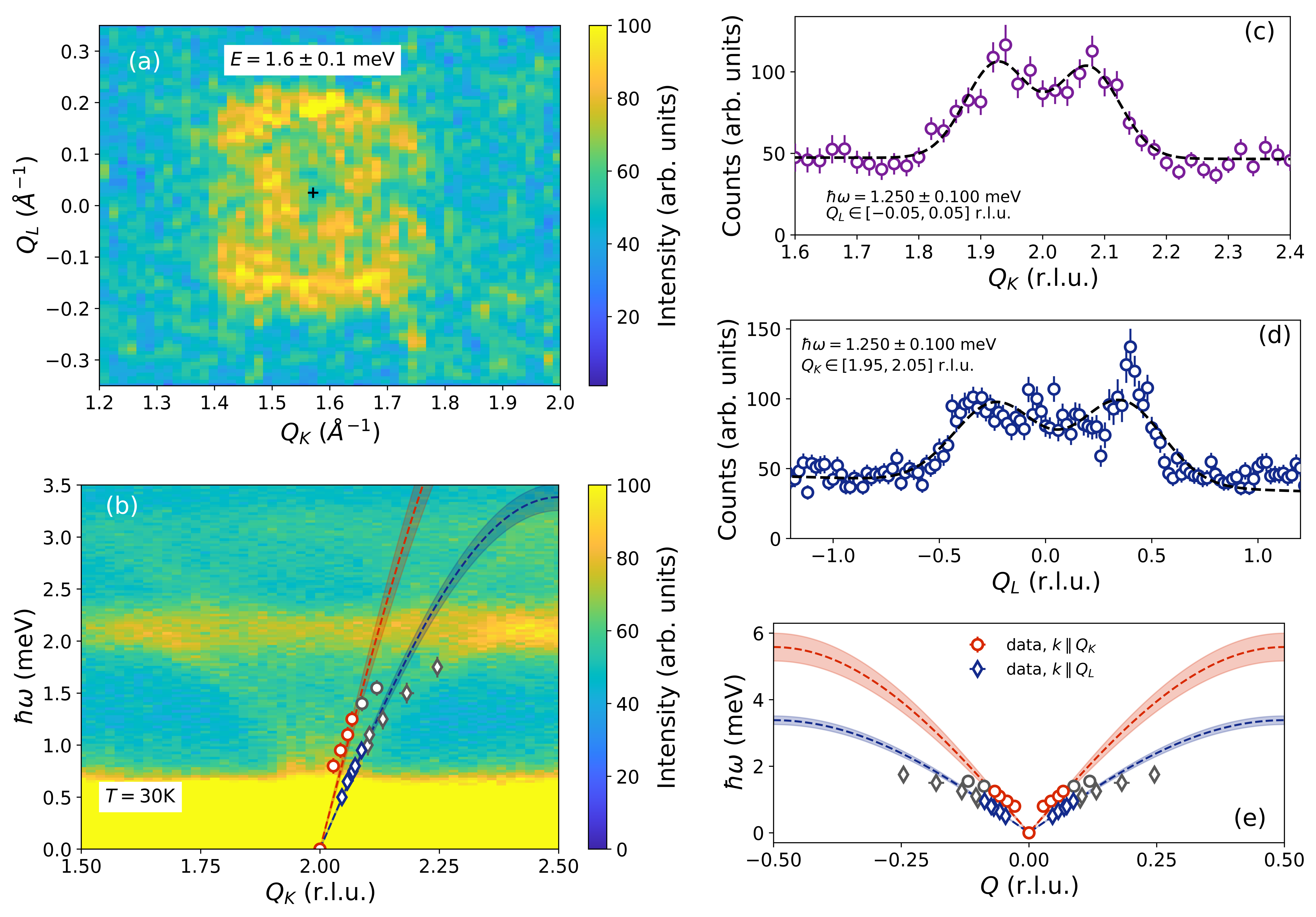}
    %\subfigure{\includegraphics[width={0.5635\textwidth}]{figures/PUB_Cu-CPA_IN5_4AA_phonon_020_2D_map_scale2.png}}
    %\subfigure{\includegraphics[width=0.4165\textwidth]{figures/CPA_4A_IN5_Qcuts_phonons_E1150_1350_QH__QK0020_3000_dQK0020_QL-050_0050_BGscale2_.pdf}}
    %\subfigure{\centering\includegraphics[width={0.5635\textwidth}]{figures/SUP__phonons_4AA_30K_scale2.png}}
    %\subfigure{\includegraphics[width=0.4165\textwidth]{figures/CPA_4A_IN5_Qcuts_phonons_E1150_1350_QH__QK1950_2050_QL-1800_1800_dQL0020_BG_scale2.pdf}}
\caption{\label{fig:phonons_4A_30K}
{\bf Low-energy phonon dispersion of Cu-CPA, measured with $E_i = 5.11$~meV at $T = 30$~K.} (a) Constant-energy slice integrated over the range (1.5, 1.7)~meV to identify two rings in ($Q_K$,$Q_L$), around the point $\mathbf{Q} = (0~2~0)$, which are approximately circular in absolute units (\AA$^{-1}$). (b) Intensity, $I(Q_K,\omega)$, obtained by integration over $Q_H$ in the range [$-0.85$, 0.85]~r.l.u.~and $Q_L \in [-2, 4]$~r.l.u.~to show the phonon contribution as a function of $Q_K$. Red circles represent phonon 2, which has stronger intensity along $Q_K$ and is shown in panel (c); blue diamonds represent phonon 1, which has stronger intensity along $Q_L$ and is shown in panel (d). \textcolor{blue}The black, dashed line correspond to the fit of the acoustic phonon, and the red, respectively blue shaded area correspond to a $1\sigma$~confidence interval. (c,d) Characteristic constant-energy cuta prepared by integration over small energy and wave-vector intervals around $Q_L = 0$ and $Q_K = 2$, which yield the intensity profiles as functions of $Q_K$ (c) and $Q_L$ (d). (e) Deduction of two phonon dispersions from multiple constant-energy cuts. Dashed lines represent the best fits to these phonons and shaded areas the $\sigma$ fitting uncertainties. Gray symbols were excluded from the velocity-fit.}
\end{figure*}

The incident neutron energies we used to study the different excitation-energy ranges were 2.27~meV ($\lambda = 6$~\AA), 2.70~meV ($\lambda = 5.5$~\AA), and 5.11~meV ($\lambda = 4$~\AA). To study the bottom of the magnetic excitation spectrum and the low-temperature acoustic phonon contribution, both at temperature $T = 0.05$~K, we used $E_i = 2.27$~meV to take advantage of the high energy resolution ($\Delta  E \approx 28$~$\mu$eV at the elastic line and even better at finite energy transfer); this high resolution and the clean signal made it unnecessary to subtract any high-temperature scattering data representing the background. To investigate the complete magnetic spectrum and the temperature-dependence of the phonon contribution, we used both $E_i = 2.70$~meV ($\Delta E \approx 35$~$\mu$eV) and $E_i = 5.11$~meV ($\Delta E \approx 85$~$\mu$eV), performing our measurements in a dilution refrigerator spanning the range 0.05--30~K. The quoted energy resolutions correspond to the half-width at half-maximum (HWHM) height of the quasielastic line.

Finally, to investigate the effects of isotope substitution on the elastic properties, we performed an additional experiment on the LET spectrometer at ISIS (Chilton, UK), using five large single crystals, four of which were non-deuterated, with a net mass of $626.6(1)$~mg, and one deuterated with mass $14.5(1)$~mg. These crystals were co-aligned within 1.0$^\circ$, with the $b$ and $c$ directions defining the scattering plane. We took advantage of the energy-multiplexing capabilities of LET to measure simultaneously with the four incident neutron energies $E_i = 1.70$, 2.40, 3.70, and 6.40~meV. Due to their intrinsic energy scales, we resolved the magnetic excitations in our datasets taken with both $E_i = 2.40$ and 3.70~meV, while the phonons were best analyzed using $E_i = 3.70$~meV, where the energy resolution was $\Delta E \simeq 50$~$\mu$eV. Our CAMEA data were analyzed using the software package MJOLNIR \cite{lass_mjolnir_2020} and our time-of-flight data using Horace \cite{ewings_horace_2016}.

\begin{figure*}[t]
    \begin{subfigure}%
        \centering
        \includegraphics[width={0.5635\textwidth}]{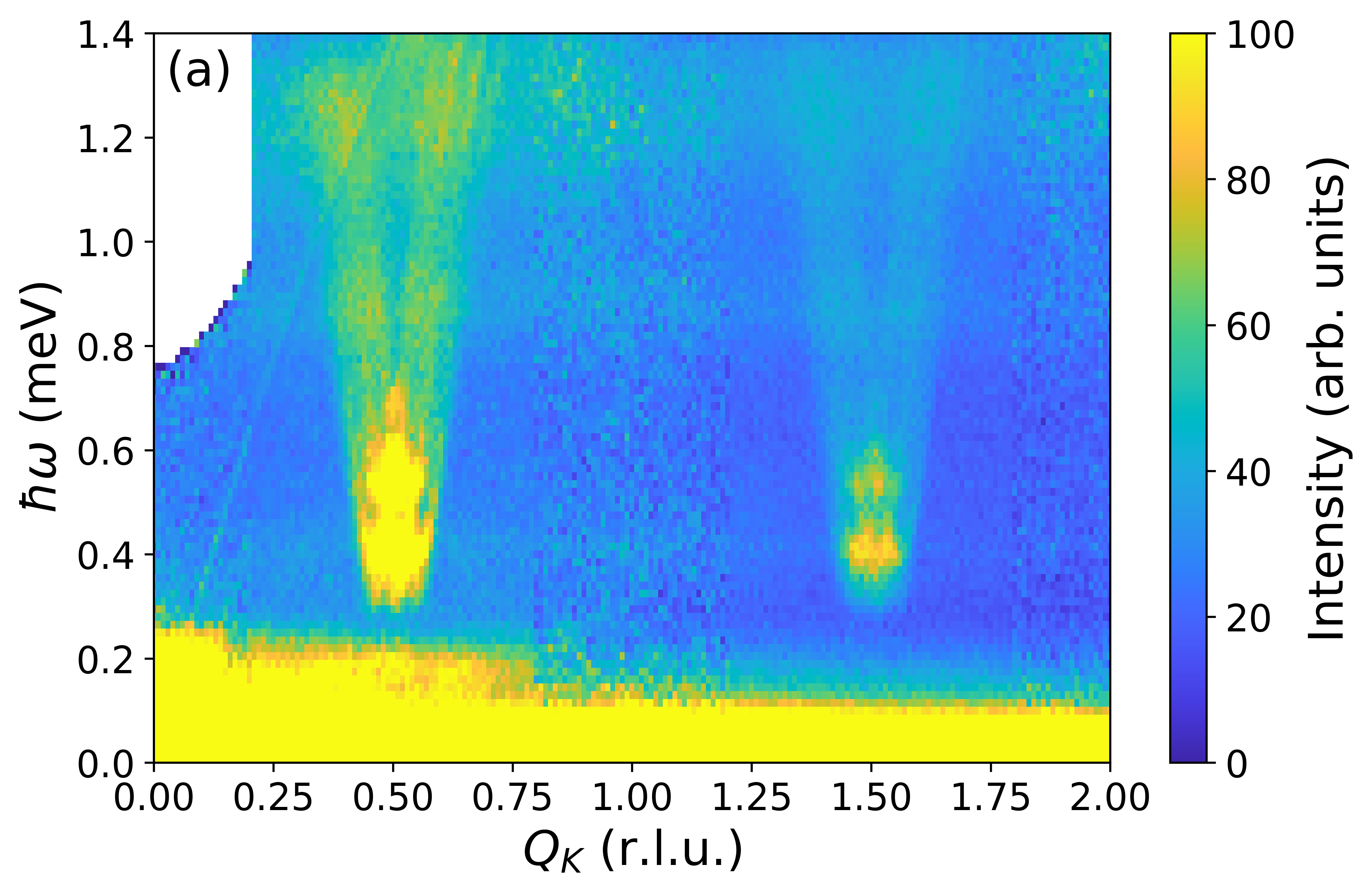}
    \end{subfigure}
    \begin{subfigure}
        \centering
        \includegraphics[width={0.4165\textwidth}]{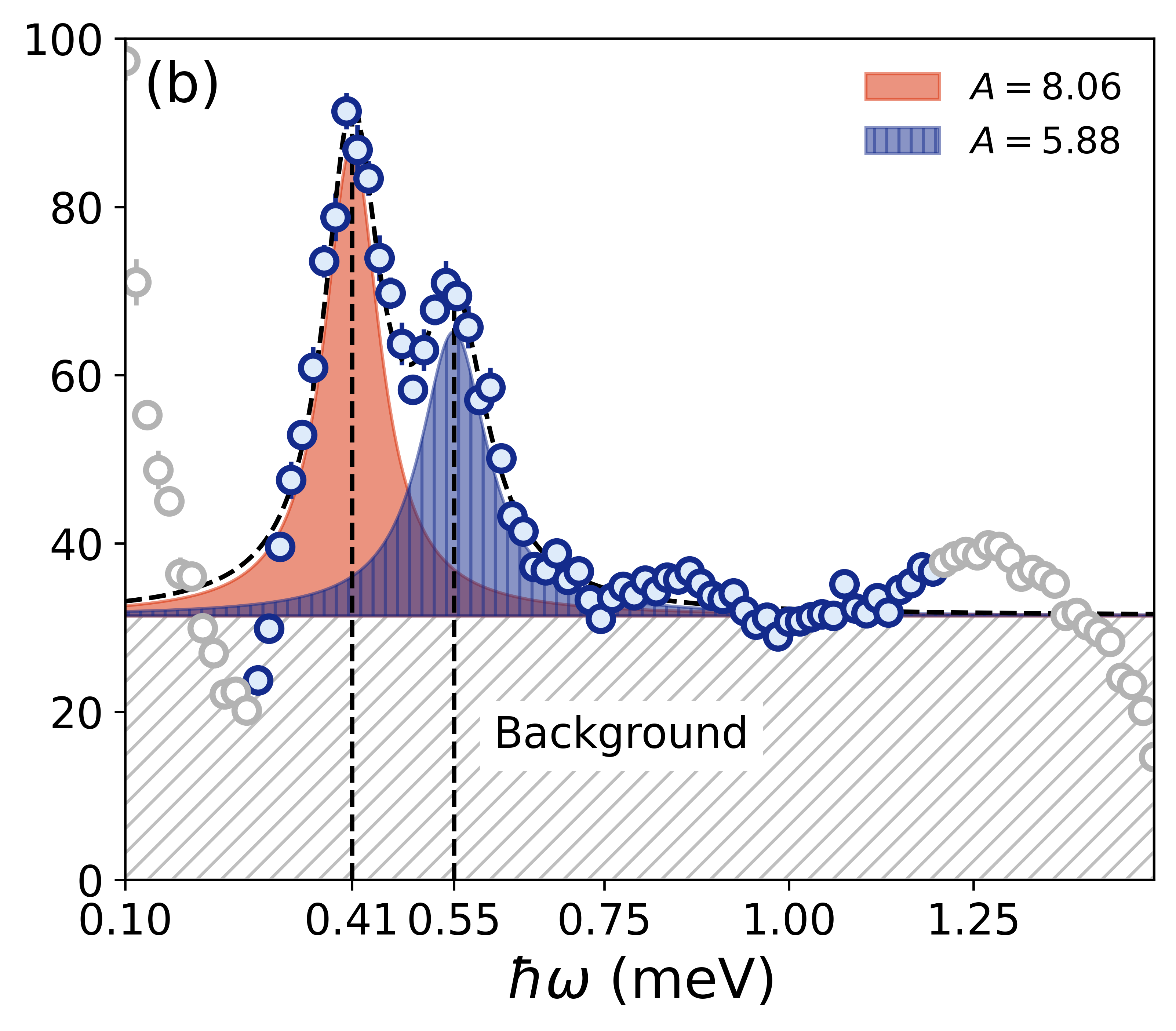}
    \end{subfigure}
\caption{\label{fig:dispersion_6A}
{\bf Low-energy triplon dispersion of Cu-CPA, measured with $E_i = 2.27$~meV at $T = 0.05$~K.} (a) Intensity, $I(Q_K,\omega)$, obtained by integration over $Q_H \in [-0.85, 0.85]$~r.l.u.~and over the full range of $Q_L$ to reveal the two dispersive triplon branches. Bragg peaks and phonon contributions have been masked. (b) Constant-{\bf Q} cut with $Q_K = 1.5$~r.l.u.~and integration width $dQ_K = \pm 0.02$~r.l.u., integrating over $Q_H \in [-0.85, 0.85]$~r.l.u.~and over the range $Q_L \in [-2, 4]$~r.l.u. The two peaks were fitted with Lorentzians, whose center positions (dashed lines) were taken to estimate the two gaps and whose integrated intensities are shown in the legend. The fitting range (blue circles) was reduced to avoid the elastic line, the extension of its tail to $\hbar\omega \simeq 0.25$~meV, and a dispersionless feature around 1.2~meV (data points not included in the fit are shown as gray circles). The hatched area represents a constant background contribution.}
\end{figure*}

\section{Spin and phonon excitation spectra}
\label{sres}

We begin with the overview of the excitation spectrum shown in Fig.~\ref{fig:excitation_spectrum_5p5AA}(a). It is evidently rather complex, with clear magnetic excitations, more than one type of phonon contribution on overlapping energy scales, and the usual instrumental backgrounds. One standard method for separating these contributions is through their $Q$-dependence, because the spin excitations follow the magnetic form factor of Cu$^{2+}$ whereas the phonons have a universal $|Q|^2$ form. In temperature, the phononic contributions can usually be isolated by measuring at higher thermal occupations, whereas the magnetic excitations in Cu-CPA vanish beyond 10~K and no magnetic contributions are discernible above 30~K. Nevertheless, a clear separation of the different components is made quite challenging in Cu-CPA by some unusually low optical phonon energies and an ``elastomagnetic'' effect of the magnetic correlations on these phonons. To achieve the required separation, we analyze first its most straightforward contributions, which are the acoustic phonon branches and then the magnetic excitation branches in Fig.~\ref{fig:excitation_spectrum_5p5AA}(a). After understanding these components, we characterize the unconventional phononic contribution with the aid of a reinterpretation of the measured specific heat.

\subsection{Acoustic phonons}

To isolate the acoustic phonon contribution to the spectrum, we investigate the excitation spectrum at high $Q$, which in Cu-CPA means around the point $\mathbf{Q} = (0~2~0)$, and at our highest temperature of $T = 30$~K, where no coherent magnetic excitations are present. A constant-energy scan at 1.6 meV, shown in Fig.~\ref{fig:phonons_4A_30K}(a), reveals a pair of concentric rings that are approximately circular, indicating the presence of two acoustic phonons. Full details of our data treatment are presented in App.~\ref{app:accoustic_phonons}. By following the intensity profiles as functions of $Q_K$ and $Q_L$, shown respectively in Figs.~\ref{fig:phonons_4A_30K}(c) and \ref{fig:phonons_4A_30K}(d) for one example energy slice, we reconstruct the dispersions of these two phonon branches to obtain the points shown in Fig.~\ref{fig:phonons_4A_30K}(e). In Fig.~\ref{fig:phonons_4A_30K}(b), where we have integrated over $Q_L$ to show the intensity $I(Q_K,w)$, we notice that both phonon branches are visible, with the weaker branch corresponding well to the overplotted $Q_L$ dispersion (blue diamonds). We remark here that both acoustic phonon branches seem to undergo an avoided crossing with the dispersionless modes providing the intense signal around 2~meV, and hence that one cannot expect follow these branches to their upper band edges. 

We proceed by assuming the dispersion relation $\hbar \omega(k) = 2 \hbar \sqrt{C/M} \sin (kd/2)$, where $C$ is an effective spring constant, $M$ an effective mass, and $d$ the unit-cell dimension in the appropriate crystallographic direction. The speed of sound in the acoustic (small-$k$) regime is then directly proportional to the ``reduced spring constant'', $\sqrt{C/M}$, and also to the unit-cell dimension, as we show in App.~\ref{app:accoustic_phonons}. Noting that the speed of sound is defined strictly at $Q = 0$, we measure an approximation to it by working over a finite $Q$ range, obtaining the two sound velocities $v_1 = 320(24)$~m$\,$s$^{-1}$ for the softer mode (visible predominantly along $Q_L$) and $v_2 = 555(42)$~m$\,$s$^{-1}$ for the stiffer one (visible along $Q_K$). Thus the velocities of sound in Cu-CPA are rather low, as expected for a soft organic compound. In App.~\ref{app:accoustic_phonons} we quantify the four reduced spring constants obtained from these velocities, 
noting that our results show Cu-CPA to be structurally stiffer along the ladder ($b$) direction, which is a stack of inorganic (rung) units, than in the transverse direction, where the ladders are separated by large organic groups [Fig.~\ref{fig:excitation_spectrum_5p5AA}(b)]. 

\subsection{Magnetic excitations}

The dynamical structure factor of the two-leg $S = 1/2$ quantum spin ladder with pure-Heisenberg rung and leg interactions is dominated by a dispersive $\Delta S_z = 1$ excitation branch known as the ``triplon.'' This magnetic quasiparticle, which can be considered as a propagating rung singlet-triplet excitation dressed by quantum fluctuations, is threefold degenerate at zero field and appears in the antisymmetric parity sector. Numerical calculations, primarily by ED and by the density-matrix renormalization-group (DMRG) or MPS techniques, have predicted the spectrum for all leg-to-rung-to-leg interaction ratios, $\alpha$ \cite{barnes1993, Johnston2000, Schmidiger2013}, and show that the spectrum at intermediate $\alpha$ exhibits features combining those of the strong-dimer and strong-chain regimes. As noted in Sec.~\ref{sintro}, the strong-rung regime ($\alpha \le 0.5$), has been studied in detail in the context of BPCB and BPCC, and without an applied magnetic field is amenable to a number of analytical or expansion methods that yield the exact spectral shape and intensity. At higher $\alpha$, however, only numerical methods are able to compute the spectrum with quantitative accuracy \cite{Schmidiger2013}. 

Although the magnetic component dominates the excitation spectrum at low temperatures and small $Q$ values, as we showed in Fig.~\ref{fig:excitation_spectrum_5p5AA}(a), a quantitative analysis requires a systematic subtraction of the phonon contribution. To deduce the phononic background to subtract from the spectrum measured at base temperature ($T = 50$~mK), we treated our high-temperature dataset ($T = 30$~K) by first masking the Bragg peaks and their tails, specifically from the elastic line up to a threshold energy of 0.2~meV. We then isolated the phonon contribution by considering only the second Brillouin zone, away from the spurious intensity contribution from the beam-stop, and finally rescaled this intensity by the Bose function to account for the measurement temperature. The non-magnetic background therefore contained a temperature-independent part (around the elastic line) and a temperature-dependent contribution most relevant at higher energy transfers, while the instrumental background was assessed by a vanadium scan.

We remark here that, in all our discussions of the magnetic excitations, we assume based on the structure of Cu-CPA that these are entirely one-dimensional in nature and have  integrated the INS intensities we analyze in both perpendicular momentum-transfer directions. We relegate to App.~\ref{app:1D} the demonstration of this 1D character from our INS measurements.

\subsubsection{Low-energy spectrum} 
\label{ssec:low_energy_spectrum}

We begin by addressing the low-energy part of the spectrum, which we measured at the extremely high energetic resolution $\Delta  E \simeq 40$~$\mu$eV available on IN5 with $E_i = 2.27$~meV. The low-energy spectrum is shown in Fig.~\ref{fig:dispersion_6A}(a), and as expected from the low-temperature crystal structure displays two distinct, gapped branches that disperse steeply from the half-integer $Q_K$ points. The intensity $I(\hbar \omega)$ measured at $Q_K = 1.5$~r.l.u.~is shown as the open circles in Fig.~\ref{fig:dispersion_6A}(b). We fitted the peaks with two Lorentzians above an assumed constant background, excluding the near-elastic contributions and non-dispersive features beyond 1.2~meV by restricting the fitting range. This procedure yielded the fitting parameters shown in Table \ref{tab:Cu-CPA_6A_best_fit}. A key result is an accurate estimate of the two gaps as $\Delta_b = 0.41(1)$ meV and $\Delta_a = 0.55(1)$ meV, a 34\% difference corresponding well with the distortion parameter $\delta = 13.5$\% estimated in Ref.~\cite{philippe_metal-organic_2024}. Here we have assigned the smaller spin gap to Ladder $b$ of Fig.~\ref{fig:excitation_spectrum_5p5AA}(a) and the larger to Ladder $a$, basing this attribution on the observation of Ref.~\cite{philippe_metal-organic_2024} that largest difference between the two inequivalent ladders is in the Br-Br separations within the rung interaction pathways; the significantly shorter distance for Ladder $a$ is likely to mean a stronger $J_{\rm rung}$ on this ladder and hence a larger gap. Because the mode linewidths, $\Gamma_a$ and $\Gamma_b$, are much larger than the instrumental resolution, we conclude that they are intrinsic. The integrated intensities of the two peaks, shown by the red and blue shading in Fig.~\ref{fig:dispersion_6A}(b), obey the relation
\begin{equation}
    \label{eq:ratio}
    r_{b/a} \! = \! \frac{S_b(Q) \hbar\omega_{\mathrm{Q},b}}{S_a(Q) \hbar\omega_{\mathrm{Q},a}} \! = \! \frac{8.06(60) \! \times \! 0.41(1)}{5.88(70) \! \times \! 0.55(1)} \! = \! 1.0(2),
\end{equation}
and hence reveal equal scattering contributions after correcting for the excitation energy.

\begin{figure*}[t]
    \includegraphics[width = \textwidth]{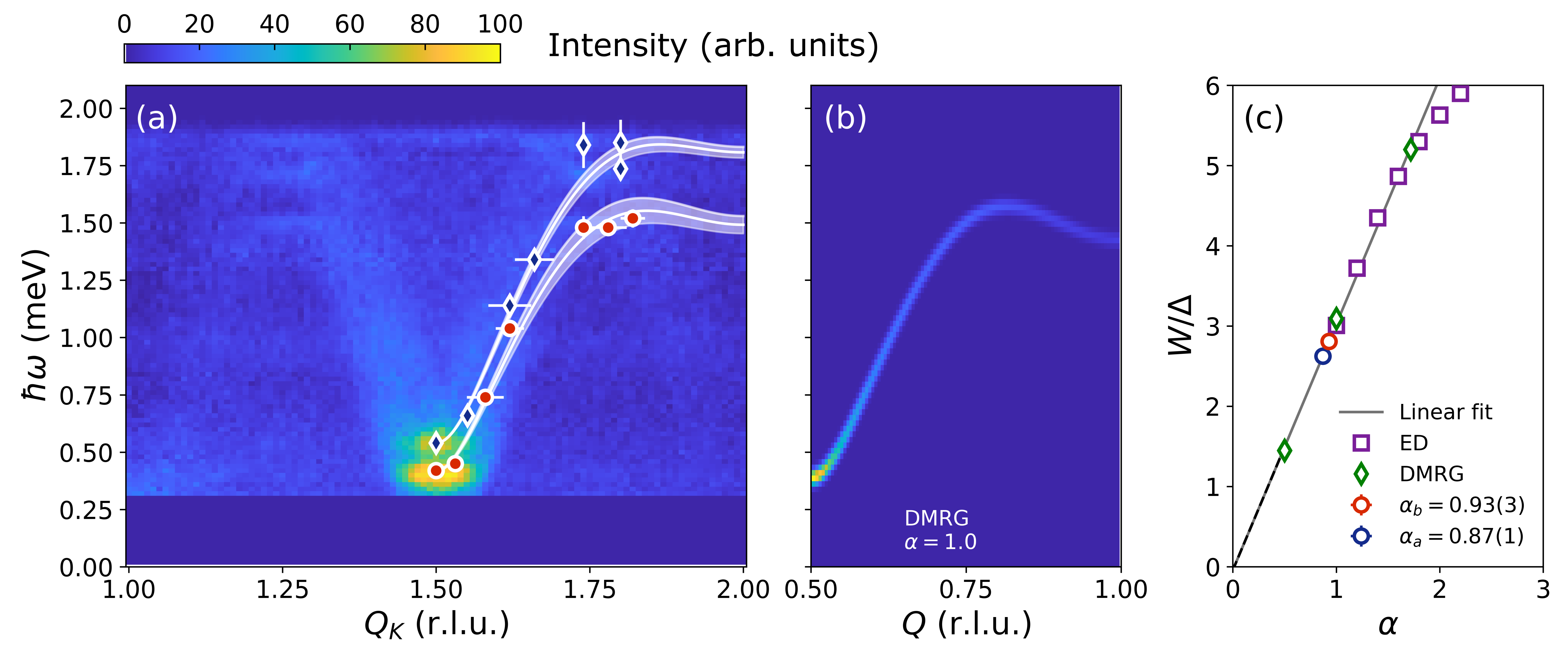}
\caption{\label{fig:full_dispersion}
{\bf Full magnetic excitation spectrum of Cu-CPA.} 
(a) Double triplon dispersion determined by subtraction of the background and of the complete phonon contribution. The measurements were performed with $E_i = 2.70$~meV. We integrated the data over the ranges $Q_H \in [-0.85, 0.85]$~r.l.u.~and $Q_L \in [-2, 4]$~r.l.u.
Red circles (Ladder $b$) and blue diamonds (Ladder $a$) show the center positions of Lorentzian (Gaussian) fits to selected constant-{\bf Q} (constant-energy) scans. Solid lines show best fits obtained by parameterizing the ladder dispersion relation as a function of $\alpha$ \cite{barnes1994}, as detailed in App.~\ref{app:sec:parametrized_SMA}.
(b) Dynamical structure factor, $S(Q,\omega)$, calculated by density-matrix renormalization-group (DMRG) methods \cite{Schmidiger2013} for a two-leg ladder with leg-to-rung interaction ratio $\alpha = 1$ and with the energy scale selected to fit the gap $\Delta_b$ measured in experiment. 
(c) Bandwidth-to-gap ratio, $W/\Delta$, of the ladder triplon, computed for a range of $\alpha$ values by ED and DMRG and compared with the results deduced for Ladders $a$ and $b$. We draw attention to the remarkably wide regime of linear dependence (up to $\alpha = 1.6$) where $W/\Delta \approx 3 \alpha$.}
\end{figure*}

\begin{table}[b]
\centering
\begin{tabular}{l|c|c}
        Parameter & $\;$Ladder $b$$\;$ & $\;$Ladder $a$ $\;$\\
        \midrule
        Energy gap ($\Delta$, meV) & $\;$0.41(1)$\;\;\;$ & $\;$ 0.55(1)$\;\;\;\;\;\;$ \\
        Linewidth ($\Gamma$, meV) $\;$ & $\;$ 0.092(7) $\;$ & $\;$ 0.110(14) $\;$ \\
        Intensity ($A$, a. u.) $\;$ & $\;$ 8.06(60) $\;$ & $\;$ 5.88(70) $\;\;$ \\
\end{tabular}
\caption{Best-fit parameters for the two Lorentzian peaks shown in Fig.~\ref{fig:dispersion_6A}(b). The energy is taken from the peak center, $\Gamma$ is the FWHM, and $A$ is the integrated intensity.}
% JP: Comment to FWHM: DGM asked whether it was the FWHM or sigma of the Lorentzian, I checked, it is the FWHM (hence correct as is). 
    \label{tab:Cu-CPA_6A_best_fit}
\end{table}

\subsubsection{Two-ladder dispersion}

Figure \ref{fig:full_dispersion}(a) shows the magnetic excitation spectrum of the two-ladder system measured with $E_i = 2.70$ meV at $T = 50$ mK. Qualitatively, we observe two dispersive branches that remain distinct at all $Q_K$, with the gaps quoted in Table \ref{tab:Cu-CPA_6A_best_fit} at half-integer $Q_K$, band maxima around $\pi/2$ ($Q_K = 0.25$ r.l.u.), and an apparent drop in mode energy towards integer $Q_K$. In Fig.~\ref{fig:full_dispersion}(b) we show that the shapes of both dispersive triplon branches, particularly that of Ladder $b$, are quite similar to that of the canonical $\alpha = 1$ ladder computed by DMRG in Ref.~\cite{Schmidiger2013}. 

The highest energies we measured near the two upper band edges, 1.48(5)~meV for the triplon branch of Ladder $b$ and 1.85(10)~meV for Ladder $a$, differ by approximately 25\%, whereas the gaps differ by 34\% (Sec.~\ref{sres}B1), indicating that triplon $a$ has a smaller bandwidth-to-gap ratio than $b$. For completeness we remark that we see no evidence for a crossing of the two triplon dispersions, but nor can we exclude such a crossing only on the basis of Fig.~\ref{fig:full_dispersion}(a). The mode intensities decrease with $Q_K$ away from the gapped points until they vanish completely at the zone boundary; although we cannot access the zone boundary, we can attempt to estimate the $Q_K$ position of the band maximum, the inflection point of the dispersion, and the bandwidth-to-gap ratio, thereby placing some constraints on viable $\alpha$ values for each triplon branch.

\begin{figure*}[t]
\centering
\subfigure{\includegraphics[width=0.48\textwidth]{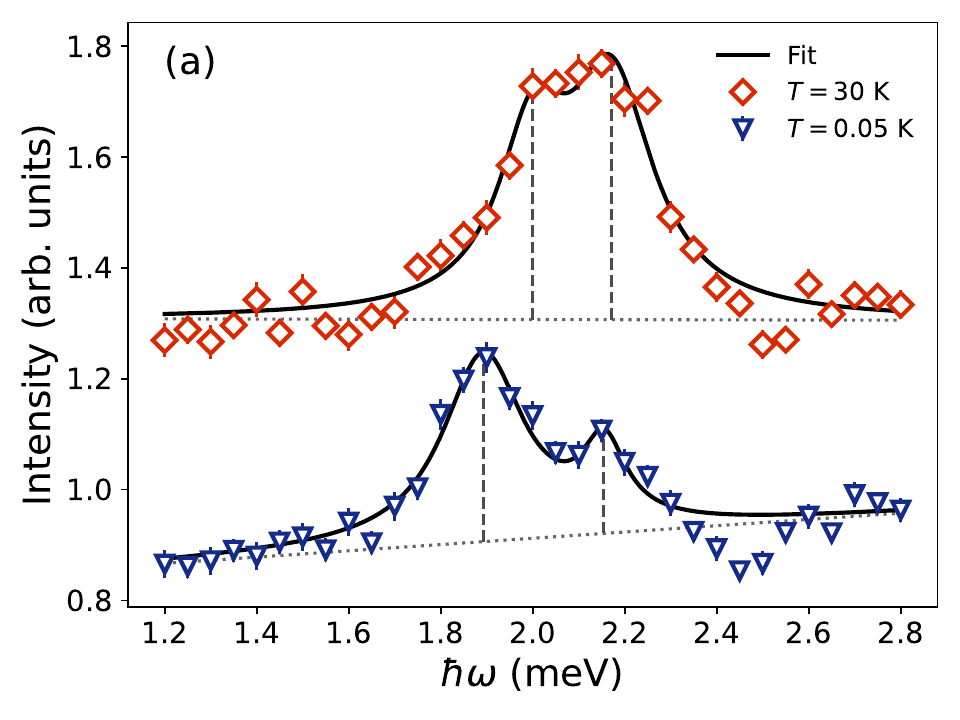}}
\subfigure{\includegraphics[width = 0.48\textwidth]{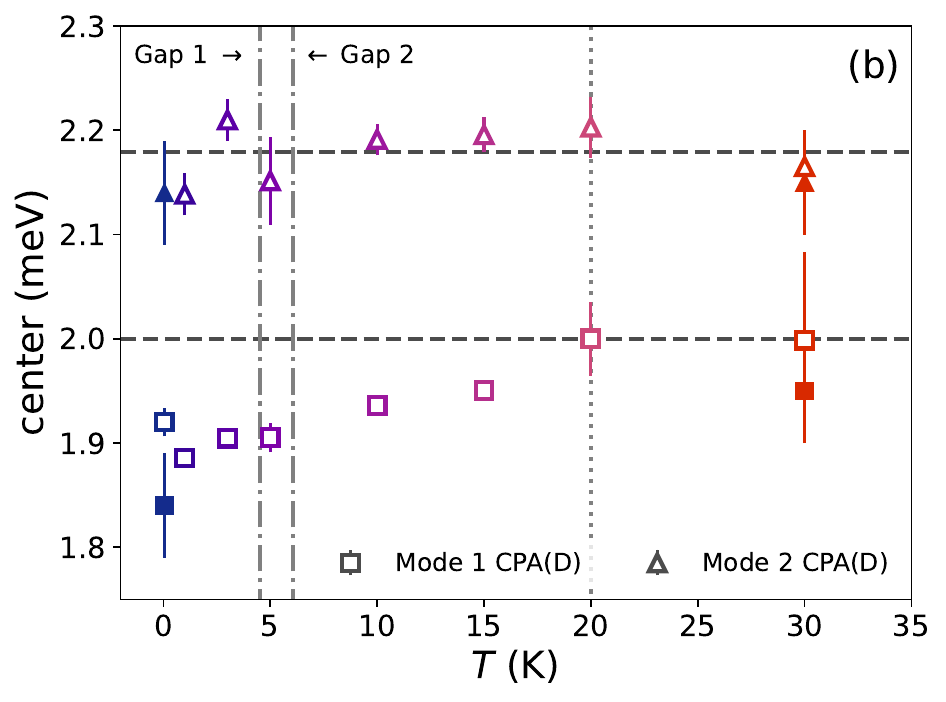}}
\caption{{\bf Elastomagnetic effects.} (a) Scattered intensity obtained by integration over the momentum ranges $Q_H \in [-0.85, 0.85]$~r.l.u., $Q_K  \in [1.88, 1.92]$~r.l.u., and $Q_L \in [-2, 4]$~r.l.u., shown as a function of energy transfer at our highest and lowest temperatures to characterize the evolution of the two distinct low-lying vibrational modes associated with the CPA molecules. The lower mode shifts to lower energies upon cooling, whereas the upper mode remains at a constant energy but appears to lose intensity. (b) Temperature-dependence of the center positions of the two vibrational modes (triangles and squares) required to fit the scattered intensity, obtained from Lorentzian profiles. The dashed lines serve as guides to the eye. In most cases, the error bars are smaller than the symbol sizes. The downward shift in the lower mode (squares) below 20~K suggests its association with the onset of magnetic correlations.}
\label{fig:elastomagnetic}
\end{figure*}

For a quantitative analysis, we perform ED calculations for a two-leg ladder with antiferromagnetic Heisenberg interactions $J_\mathrm{leg}$ and $J_\mathrm{rung}$ in the ratio $\alpha$. In Fig.~\ref{fig:full_dispersion}(c) we compare our results for the bandwidth-to-gap ratio as a function of $\alpha$ with those obtained by DMRG calculations for different $\alpha$ values in Ref.~\cite{Schmidiger2013}. It is clear that the bandwidth-to-gap ratio, which in principle is an experimental observable, constitutes an excellent measure of the $\alpha$ parameter of a ladder, with a monotonic dependence that is even linear over a remarkably broad range of $\alpha$. Because the bandwidth is more sensitive to $J_\mathrm{leg}$ and the gap to $J_\mathrm{rung}$, the monotonic dependence is no surprise, but the linearity up to $\alpha \approx 1.6$, where multiple nonlinear effects become apparent in any strong-coupling expansion, is surprising. 

Because it is difficult to estimate the upper band edges accurately from our INS data [Fig.~\ref{fig:full_dispersion}(a)], we proceed by applying the parametrization of Ref.~\cite{barnes1994} to the triplon dispersions computed by DMRG in Ref.~\cite{Schmidiger2013M}. Full details of the parametrization procedure are presented in App.~\ref{app:sec:parametrized_SMA}. By this approach we obtain a quantitative estimate of the interaction parameters of the two ladders as $J_{\mathrm{leg},b} = 0.74(3)$~meV and $J_{\mathrm{rung},b} = 0.80(4)$~meV [$\alpha_b = 0.93(3)$] for the lower branch and $J_{\mathrm{leg},a} = 0.84(2)$~meV and $J_{\mathrm{rung},a} = 0.97(2)$~meV [$\alpha_a = 0.87(1)$] for the upper. Thus, we verify the deductions (above) that both ladders have $\alpha$ close to 1 and that $\alpha_a < \alpha_b$.

\subsection{Spin-sensitive phonon modes}

The final contribution to the spectrum that remains present after subtraction of the acoustic phonons [Fig.~\ref{fig:phonons_4A_30K}(a)] and the triplons [Fig.~\ref{fig:full_dispersion}(a)] is a flat, non-dispersive feature observed around 2.0~meV at $T = 30$~K [Fig.~\ref{fig:phonons_4A_30K}(c)]. In App.~\ref{app:cpa_mass_vibration_mode} we show that the powder average of this intensity is proportional to $|Q|^2$, verifying its phononic nature. We then demonstrate that the energy of this feature has an H-D isotope effect, making a clear case for assigning it to a vibrational mode or modes of CPA molecules. More specifically, in App.~\ref{app:cpa_mass_vibration_mode} we show that the average vibrational energy is fully consistent with a simple proportionality to $1/\sqrt{m}$, where $m$ denotes the mass of CPA molecules with no, partial, or full deuteration.

Analysis of the integrated scattering intensity as a function of energy reveals that the flat band is composed of two distinct modes, seen in Fig.~\ref{fig:elastomagnetic}(a), whose average positions evolve partially with temperature. Qualitatively, this temperature dependence is isotope-independent, and in Fig.~\ref{fig:elastomagnetic}(b) we concentrate on deuterated Cu-CPA. The upper of the two modes is temperature-independent, and is centered at $\hbar\omega = 2.17(3)$~meV, while the energy of the lower is reduced upon cooling, from $\hbar\omega  = 2.00(2)$~meV at 30~K to 1.90(2)~meV below 10~K. This temperature range coincides with the emergence of magnetic correlations, which become complete below the gap temperature, and hence corroborate an elastomagnetic effect. 

\begin{figure*}
    \centering
    \includegraphics[width = \textwidth]{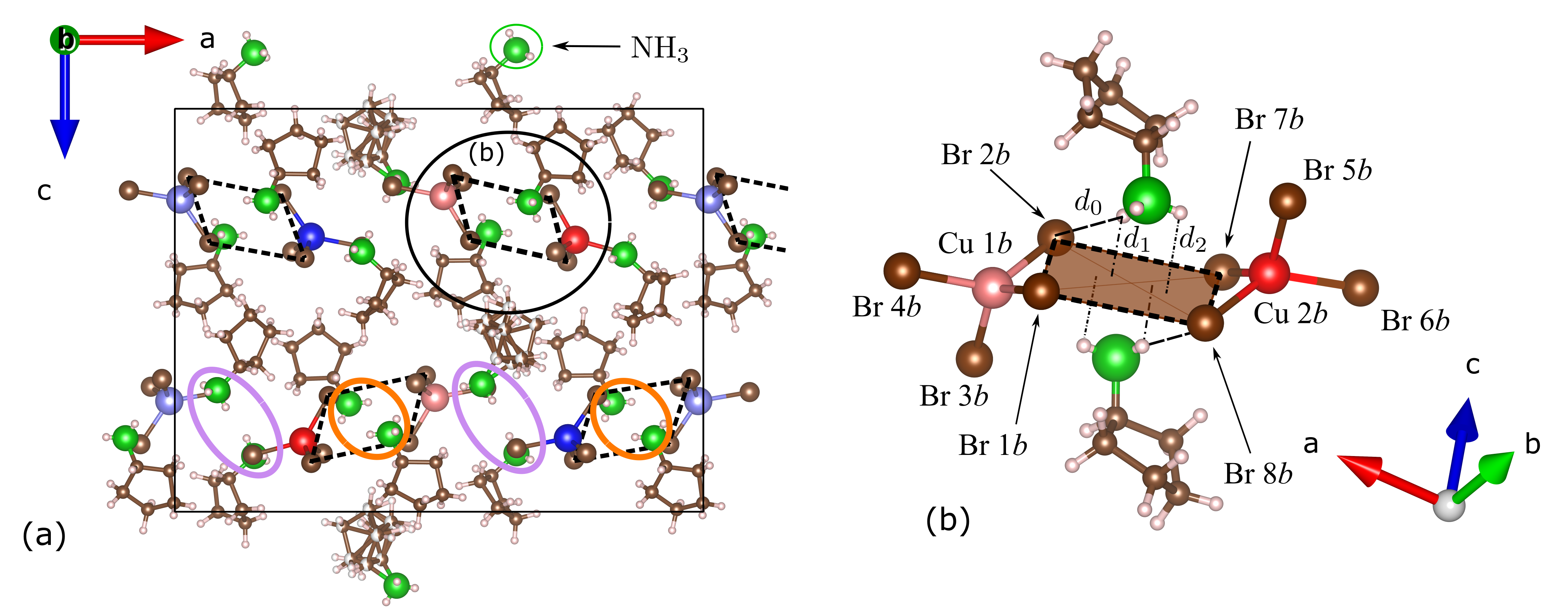}
    \caption{{\bf Involvement of CPA molecules in ladder magnetism}. (a) Low-temperature crystal structure of Cu-CPA ($T = 85$~K), showing four ladder rungs that each contain a pair of NH$_3$ radicals from a CPA molecule within the rung-superexchange volume; two of these pairs are circled in orange. In contrast, the NH$_3$ radicals of the remaining two CPA molecules per ladder are situated well outside the rungs; two of these pairs are circled in purple. The black circle denotes one rung unit that is displayed in a different projection in panel (b). (b) Visualization of the rung superexchange volume based on the nearly-square trapezium formed by the four Br atoms involved in the two Cu-Br-Br-Cu paths. The proximity of individual H atoms of the ammonium radical to the nearest Br atom ($d_0$) and to the rung plane ($d_1$ and $d_2$) suggests the involvement of ammonium electrons in virtual hopping processes between the halogen atoms, and hence in the rung spin interaction.}
    \label{fig:crystal_structure_NH3}
\end{figure*}

This change in stiffness of the lower dispersionless phonon mode correlates closely with changes in the local susceptibility that occur as magnetic correlations set in. The lower local phonon undergoes a 5\% softening as a result of this effect, which is a very large relative change when compared with the 0.5\% effect observed in the compound (C$_4$H$_{12}$N$_2$)Cu$_2$Cl$_6$~(PHCC) \cite{Bettler2017} and the 1\% effect found in SrCu$_2$(BO$_3$)$_2$ (SCBO) under pressure \cite{Bettler2020}. To understand the origin of this ``giant'' elastomagnetic softening, we note that the four inequivalent CPA rings in Cu-CPA form two groups, one pair lying far from the Cu ladders and the other pair having their ammonium group lying very close to the ladder rungs, as shown in Fig.~\ref{fig:crystal_structure_NH3}(a).

To gauge the distances between the ammonium groups and the Br atoms lying directly on the two Cu-Br-Br-Cu superexchange pathways making up each rung, we define a plane formed by these four Br atoms, shown by the shaded surface in Fig.~\ref{fig:crystal_structure_NH3}(b), which is crudely a rhombus with an 80$^\circ$ interior angle and side lengths of 3.6~\AA~(adjoining one Cu atom) and 4.4 or 4.5~\AA~(across the rung). In both NH$_3$ radicals of a single rung unit, one of the H atoms is far (3.7~\AA) from the plane and one is located at $d_2 \simeq 3.3$~\AA~from the plane, but the third H atom is only $d_1 = 2.3$~\AA~away from the rung plane. The distance of this atom from the nearest Br atom is only $d_0 = 2.61$ \AA~[Fig.~\ref{fig:crystal_structure_NH3}(b)], which is short compared to the Br-Br spacings. Taking this H atom to measure the extent of the relevant lone-pair orbital of the strongly electronegative N atom, it is clear that this does lie close enough to affect the electronic correlations on the rung, and hence that the H atom should be considered as part of one pathway for virtual hopping processes contributing to the rung magnetic correlations. 

It then seems logical to deduce that, as magnetic correlations become established with decreasing temperature, the electronic orbitals of the H and N atoms undergo nontrivial alterations that can affect the elastic properties of the entire CPA molecule. Specifically, these changes are manifest in a substantial (5\%) drop in the stiffness of one of the local phonon modes of the CPA structure, which in turn are inherently low-lying due to their proximity to the structural phase transition occurring just above 100 K \cite{philippe_metal-organic_2024}. 

\begin{figure}
    \centering
    \includegraphics[width=0.5\textwidth]{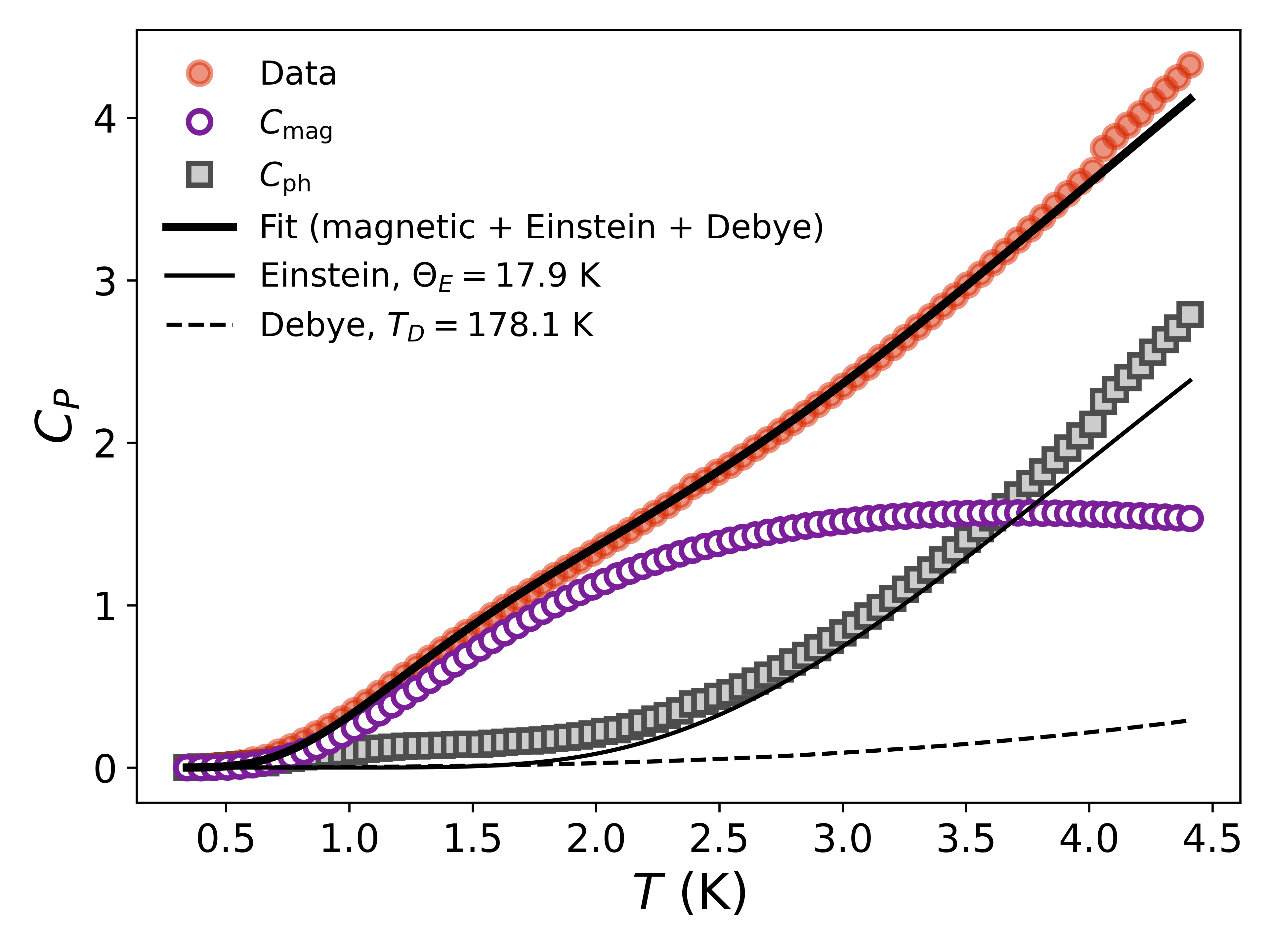}
    \caption{{\bf Low-temperature specific heat.} Three-component fit to the measured specific heat \cite{philippe_metal-organic_2024} using the precisely determined magnetic interactions, a Debye model for the phononic contribution at high temperatures, and an additional Einstein oscillator to account for the low-lying local phonon mode found by spectroscopy.}
    \label{fig:sh}
\end{figure}

\section{Elastic and magnetic contributions to the low-temperature specific heat}
\label{ssh}

In Sec.~\ref{sres} we determined the definitive magnetic interaction parameters of Cu-CPA and discovered two local (non-dispersive) phonons of the CPA molecules that are centered around 1.9~meV at low temperatures. Both results point towards the need to revisit the interpretation of the specific-heat data presented in Ref.~\cite{philippe_metal-organic_2024}. First, our determination of the four $J$ parameters leads necessarily to an unambiguous calculation of the magnetic contribution. Second, modeling the phonon contribution with the Debye formula, using a Debye temperature $T_D = 173(4)$~K chosen to fit the high-temperature specific heat ($40 \le T \le 80$~K), makes this contribution vanishingly small at low temperatures, and in no way reflective of one or more low-lying local vibrational modes. To compensate for this lack of low-energy phonon contributions, the previous best fit to a two-ladder model found the gaps $\Delta_1 = 0.32(2)$~meV and $\Delta_2 = 0.42(2)$~meV, which our INS measurements demonstrate to be artificially small.

To reconcile our specific-heat and spectroscopic measurements, we introduce one low-lying and dispersionless phonon that we assume can be modeled as a harmonic oscillator (an Einstein phonon) with an Einstein temperature $\Theta_E \simeq 20$~K (corresponding to the mode center). For the magnetic specific heat, we extend the model of Ref.~\cite{Hong2010} to the two-ladder system by fixing the two gap values to $\Delta_b = 0.41(1)$~meV and $\Delta_a = 0.55(1)$~meV, and the two magnon velocities, calculated in detail in App.~\ref{sec:magnon_velocity}, to $c_b = 1.20(5)$~meV and $c_a = 1.37(5)$~meV. 
We also retained a Debye model to account for the high-temperature contributions. By a recursive fitting of the Einstein and Debye models, in which we performed successive relaxations of the Einstein temperature and the weight factor $C_E$ of the local-phonon contribution, and then of the Debye temperature and weight factor, we obtained the values $\Theta_{\rm E} = 17.9(3)$~K, $C_{\rm E} = 8.1(3)$~JK${}^{-1}$mol${}^{-1}$, $T_{\rm D} = 178.1(44)$~K, and $C_{\rm D} = 82(2)$~JK${}^{-1}$mol${}^{-1}$. 

The fit to the data reported in Ref.~\cite{philippe_metal-organic_2024} is shown in Fig.~\ref{fig:sh} and a fit over the full temperature range can be found in App.~\ref{app:sh}. First we remark that a local phonon at $\Theta_{\rm E} = 17.9$ K, combined with the true gaps of the triplon contribution, provides a quantitatively accurate fit to the low-temperature specific heat. The weight of this local phonon is approximately 10\% of the total phonon contribution, which is an eminently reasonable value. The small deviation of the measured phononic contribution, $C_{\rm ph} (T)$, from the model over the range from 1-2 K remains as an experimental challenge. The overall fit also shows a weakness in the range 20-30 K, which could possibly be improved by adding another local phonon with $\Theta_{\rm E} \simeq 40$ K, but in the absence of spectroscopic confirmation we do not follow this empirical route. In summary, our quantitative understanding of the low-energy magnetic and phononic excitations (Sec.~\ref{sres}) leads us to an accurate determination of the specific heat, thereby unifying spectroscopy and thermodynamics. 

\section{Discussion}
\label{sdis}

By measuring the dynamical spectral function we have established that Cu-CPA possesses unusually many and low-lying optical phonon excitations. The number of phonon modes can be traced to the presence of many CPA rings per unit cell of this metal-organic material, and their nature is already suggested by the fact that an order-disorder transition between distinct CPA ring conformations was discovered at a temperature $T^* = 136$ K ($T^* = 132$ K in the deuterated system) \cite{philippe_metal-organic_2024}. As to their anomalously low-lying nature, in Ref.~\cite{philippe_metal-organic_2024} we also discovered an orthorhombic-to-monoclinic structural phase transition at $T^{\rm mono} = 113$ K ($T^{\rm mono} = 119$ K when deuterated), which suggests the presence of nearly soft phonons, potentially of several different types, in proximity to the transition. The low energies of the CPA phonons provide an additional reason for their sensitivity to the development of strong magnetic correlations on the double Cu-Br-Br-Cu superexchange paths of the ladder rungs, within which lie two ammonium groups of separate CPA molecules. A microscopic mechanism for this effect would be a target for cluster calculations addressing the spin-polarized electronic densities with and without magnetic correlations. This would then inform a discussion of the elastic properties of CPA with and without ``magnetically pinned'' ammonium.  

Turning to the magnetic excitations, we have established the presence of two conventional triplon branches whose dispersions reflect two ladders with leg-to-rung coupling ratios of $\alpha \approx 1$. We comment that the improved understanding of Cu-CPA which we have developed over the course of our studies has involved a significant downward estimation of $\alpha$, from values of order 2 based on early susceptibility measurements \cite{willett_structure_2004} to a much larger gap (smaller $\alpha$) from specific-heat measurements \cite{philippe_metal-organic_2024} to a still smaller $\alpha$ using spectroscopic information. This evolution replicates the trajectory followed by DIMPY \cite{shapiro_synthesis_2007,Hong2010,Schmidiger2011}, underlining the difficulties inherent to estimating magnetic interaction parameters on the basis of thermodynamic measurements alone. In Cu-CPA we have not been able to find any evidence for sizeable interactions between the two inequivalent ladders, which would be one route to nontrivial triplon mixing phenomena in a two-ladder system. 

Having identified both the low-energy phonons and the triplons, we have been able to provide an accurate interpretation of the low-temperature specific heat measured in Ref.~\cite{philippe_metal-organic_2024}. The sizes of the different contributions from these modes to any thermodynamic quantity remain difficult to deduce from spectroscopy, and the measurements show that approximately 10\% of the phononic contribution in Cu-CPA lies in optical phonons around 20 K (2 meV). One way to discern whether or not further local phonons are present in the spectrum beyond the range of our INS measurements would be to subtract the magnetic and Debye (high-temperature phononic) contributions from the measured specific heat, which might reveal a local-phonon signal that contains further characteristic energies above 2 meV. 

The fact that Cu-CPA contains phonon modes whose energy is very sensitive to spin correlations makes it a prime candidate for the study of hybridized spin and lattice dynamics. Taking the term magnetoelastic to
cover changes in magnetic properties due to lattice effects, we have named the converse type of behavior  we observe (changes in elastic properties due to intrinsic dynamic magnetic effects) “elastomagnetic”, where we remark that neither magnetic order nor an applied magnetic field is required to change the elastic response. Certainly a soft material such as Cu-CPA is clearly a suitable candidate for accurate measurements of magnetoeleastic behavior using both hydrostatic and uniaxial applied pressures. Dynamical spin-lattice hybridization phenomena induced by using coherent THz light to drive selected phonon modes have been given the name “magnetophononics” \cite{fechn18,giorg23,demaz25}. Although Cu-CPA can be expected to exhibit some clear magnetophononic behavior, the very low excitation energy scales and the complexity of the phonon spectrum would pose a challenge to quantitative interpretation.

To summarize, we have measured the dynamical structure factor of a previously unexplored two-ladder quantum spin system. Beyond our definitive determination of the magnetic interaction parameters of the two inequivalent ladders, we have observed low-energy spectral weight reflecting the presence of unusually low-lying and local optical phonon excitations. In addition to the expected sensitivity of the magnetic Hamiltonian to static lattice effects, we establish the sensitivity of half of the low-energy local phonons to dynamical spin correlations, in the form of appreciable shifts in the phonon energy as magnetic effects become established at low temperatures. These results position Cu-CPA as a model elastomagnetic system for future studies both by static lattice control using applied pressure and by dynamical lattice control in ultrafast nonequilibrium settings. 

\begin{acknowledgments}
We thank S. Galeski, A. Razpopov, R. Valentí, C. Kollath, B. Lake, and K. Povarov for helpful discussions. JP acknowledges the financial support of the Universit\"at Z\"urich through a UZH Candoc Grant. This project was supported by the Horizon 2020 research and innovation program of the European Union under the Marie Skłodowska-Curie Grant (Agreement No.~884104, PSI-FELLOW-III-3i). We further acknowledge the funding from the Chalmers X-ray and Neutron Initiative and support by the Wallenberg foundation though the grant 2021.0150. We are grateful to the Paul Scherrer Institute for the allocation of neutron beam-time on CAMEA at SINQ (under Project No.~20212923). Experiments at the ISIS Neutron and Muon Source were supported by beamtime allocation RB2210245 from the Science and Technology Facilities Council.
%We acknowledge in addition the support of the Chalmers X-Ray and Neutron Initiatives (CHANS) and the Swedish Research Council (VR) through a Starting Grant (No.~Dnr.~2017-05078), through Grant No.~2022-06217, through the Foundation Blanceflor 2023 fellow scholarship, and through the Area of Advance-Material Sciences of Chalmers University of Technology. 
\end{acknowledgments}

\begin{figure*}
    \centering
    %\subfigure{\includegraphics[height=5cm]{figures/CPA_fig.jpg}}
    %\subfigure{\includegraphics[height=5cm]{figures/Cu-CPA_clipped_IN5.png}}
    \includegraphics[width=0.7\textwidth]{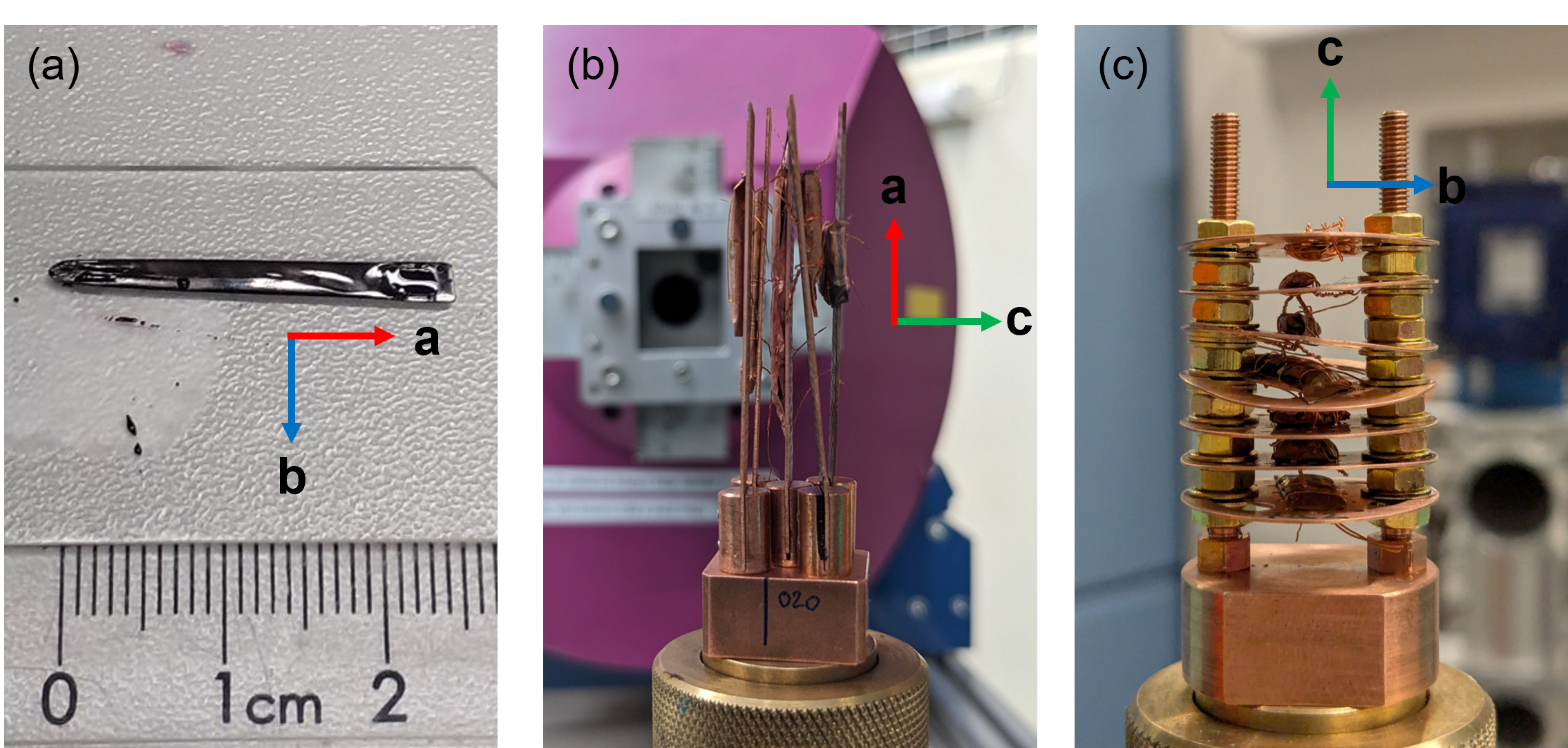}
    \caption{{\bf Samples.} Images of (a) a single crystal of deuterated Cu-CPA, (b) an assembly of five deuterated single crystals used in the IN5 experiment with $bc$ as the scattering plane, and (c) an assembly of seven single crystals used in the IN5 experiments with $ab$ as the scattering plane.}
    \label{fig:sample}
\end{figure*}

\section*{Data Availability}

The INS data on which this article is based are available in neutron facility repositories as Refs.~\cite{IN5_doi_ladder,IN5_doi_symmetric,LET_doi_ladder}.

\appendix

\section{Single crystals and INS samples}
\label{app:sample}

Figure \ref{fig:sample}(a) shows a photograph of one deuterated single crystal of Cu-CPA, as obtained from our optimized growth procedure. Figures \ref{fig:sample}(b,c) show two different assemblies of these crystals with different scattering planes, both of which were used for our neutron spectroscopy measurements at IN5.

\section{Acoustic phonon velocities}
\label{app:accoustic_phonons}

Figure~\ref{fig:excitation_spectrum_5p5AA} shows that the acoustic phonons overlap with the energy scale of the magnetic excitations and the top of the triplon bands coincides in energy with a local phonon. Extracting the magnetic excitations therefore requires a comprehensive understanding of the phononic contribution to the spectral function of Cu-CPA, and for this it is most convenient to work around $\mathbf{Q} = (0,2,0)$, where the phononic part is most intense. We reiterate that the full phonon spectrum is inordinately complicated due to the fact that there are 328 atoms in the unit cell of the low-temperature structure of Cu-CPA \cite{philippe_metal-organic_2024}. However, our previous measurements of the magnetic specific heat and susceptibility \cite{philippe_metal-organic_2024}, combined with the temperature-dependence of the magnetic excitations observed here, demonstrate that all magnetic contributions are negligible above 30~K. Thus we characterize the phononic contribution at this temperature.

Working in the $(Q_K,Q_L)$ scattering plane [Fig.~\ref{fig:phonons_4A_30K}(a)], we integrated the measured intensities over a finite energy-transfer window, $\hbar(\omega \pm \delta\omega)$, and over a narrow stripe around one of the in-plane momenta, i.e.~$\mathbf{Q} = (0, 2 \pm 0.05, Q_L)$ and $\mathbf{Q} = (0, Q_K, \pm 0.05)$, to resolve the two phonons appearing around $\mathbf{Q} = (0,2,0)$. 
All of these $Q_K$ or $Q_L$ cuts were prepared by integrating the direction parallel to the detector ($Q_H$) over the range [$-0.85$, 0.85]~r.l.u.~and the perpendicular direction over a smaller range (typically 0.1~r.l.u.). The energy-integration range, $\delta\omega$, was determined from the energy resolution (defined by the HWHM of the elastic line), which depended on the neutron wavelength. The resulting intensities, $I(Q,\omega)$, were then fitted using two Gaussians on top of a linear background, as shown in Figs.~\ref{fig:phonons_4A_30K}(c) and \ref{fig:phonons_4A_30K}(d), and the acoustic phonon dispersions gathered in Fig.~\ref{fig:phonons_4A_30K}(e).

Using the acoustic phonon dispersion relation
\begin{equation}
    \label{eq:cpa_phonons}
    \omega (k) = 2\sqrt{\frac{C}{M}}\left|\sin{\left( {\textstyle \frac12} kd \right)}\right|
\end{equation}
in the long-wavelength (small-$k$) limit yields the phonon velocity 
\begin{equation}
    \label{eq:pv}
    v = d \frac{\pi}{\hbar}\sqrt{\frac{C}{M}}.
\end{equation}
As noted in Sec.~\ref{sres}A, we found only two different sound  velocities due to the approximate fourfold symmetry of the phonon dispersion shown in Fig.~\ref{fig:phonons_4A_30K}(a), and measured these as $v_\mathrm{1} = 320(24)$~m$\,$s$^{-1}$ and $v_\mathrm{2} \simeq 555(42)$~m$\,$s$^{-1}$. Dividing by the lattice dimension and the other constants leads to the four reduced spring constants 
$\sqrt{C_{1b}/M} \simeq 1.64(13)$~meV, $\sqrt{C_{2b}/M} \simeq 2.84(22)$~meV, $\sqrt{C_{1c}/M} \simeq 0.725(55)$~meV, and $\sqrt{C_{2c}/M} \simeq 1.25(10)$~meV. This higher stiffness along the crystallographic $b$ direction is consistent with the general understanding of the crystal structure [Fig.~\ref{fig:excitation_spectrum_5p5AA}(b)], where the two pairs of inequivalent spin ladders are aligned along $b$ and separated along the $a$ and $c$ directions by the large CPA molecules. 

\begin{figure*}[t]
    \centering
    \includegraphics[width = \textwidth]{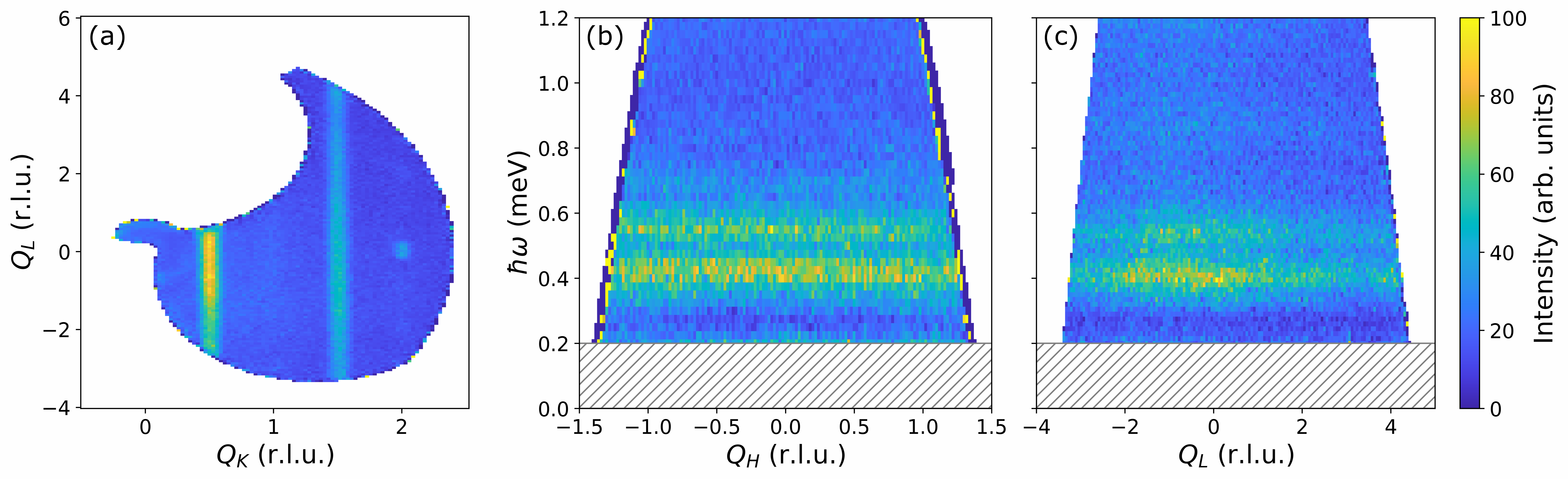}
\caption{{\bf One-dimensional magnetic behavior of Cu-CPA.} (a) Intensity contour map in $(Q_K,Q_L)$, with integration over the energy range $E = 0.4 \pm 0.1$~meV. (b) Intensity contour map in $(Q_H,E)$, with integration over $Q_K = 1.50 \pm 0.05$~r.l.u.~and $Q_L = 0.0 \pm 1$~r.l.u. (c) Intensity contour map in $(Q_L,E)$, with integration over $Q_K = 1.50 \pm 0.05$~r.l.u.~and $Q_H = 0.0 \pm 0.85$~r.l.u. The high-intensity bands shown in all three panels represent the gap(s) of the magnetic excitations.}
\label{fig:1D_nature}
\end{figure*}

\section{Triplon velocity \label{sec:magnon_velocity}}

The effective velocity of a gapped triplon branch is obtained from a parabolic approximation to the low-energy part of its dispersion \cite{Haldane_1983_nonlinear_field_theory, interacting_electrons_1994},
\begin{equation}
    \label{eq:cpa_parabolic_approximation}
    \hbar \omega_q \simeq \sqrt{\Delta^2 + c^2 |q_\parallel - \pi|^2},
\end{equation}
where $\Delta$ is the spin gap and $q_\parallel = 2\pi Q_K$. The velocity is then obtained as~\cite{Zheludev2007_dynamics_spin_liquids}
\begin{equation}
    \label{eq:cpa_magnon_velocity}
    c = \left. \frac{\mathrm{d}\sqrt{\left(\hbar\omega_q\right)^2 - \Delta^2}}{\mathrm{d}(2\pi Q_K)}\right\rvert_{Q_K = 0.5},
\end{equation}
where we computed the derivative as a divided difference, working over a relatively small range of $Q_K$ (0.06 r.l.u.) and with a very small step size of $\Delta Q_K \simeq 0.01$~r.l.u. In this way we obtained the two triplon velocities as $c_b = 1.20(5)$~meV and $c_a = 1.37(5)$~meV.

\section{1D nature of magnetic excitations \label{app:1D}}

We present INS data demonstrating that the magnetic excitations (triplons) of the two inequivalent ladders are entirely 1D in character. For this we show in Fig.~\ref{fig:1D_nature} scattered intensities measured in reciprocal-space directions perpendicular to the ladder dispersion ($Q_K$). In Fig.~\ref{fig:1D_nature}(a) we integrate over an energy range covering the gaps of both triplons, $E = [0.3, 0.7]$~meV, as well as a relatively large portion of the detector coverage in $Q_H$ ($Q_H = [- 0.75, 0.75]$~r.l.u.). This dataset was prepared using equidistant $Q$-steps of size $0.02$~\AA${}^{-1}$, which correspond to $dQ_K = 0.025$~r.l.u.~and $dQ_L = 0.056$~r.l.u. The intensities of the two clear stripes at $Q_K = 0.5$~and~1.5 follow the magnetic form factor of Cu$^{2+}$. In Fig.~\ref{fig:1D_nature}(b) we demonstrate the absence of triplon dispersion in $Q_H$ (prepared using steps of 0.075~r.l.u.) and in Fig.~\ref{fig:1D_nature}(c) the absence of dispersion in $Q_L$ (with steps of 0.056~r.l.u.); the widths of the energy bins in both panels is $\delta\hbar\omega = 15$~$\mu$eV, which corresponds to the energy resolution, and in both we integrated over $Q_K = [1.45,1.55]$~r.l.u.

\begin{figure}
    \centering
    \includegraphics[width=0.5\textwidth]{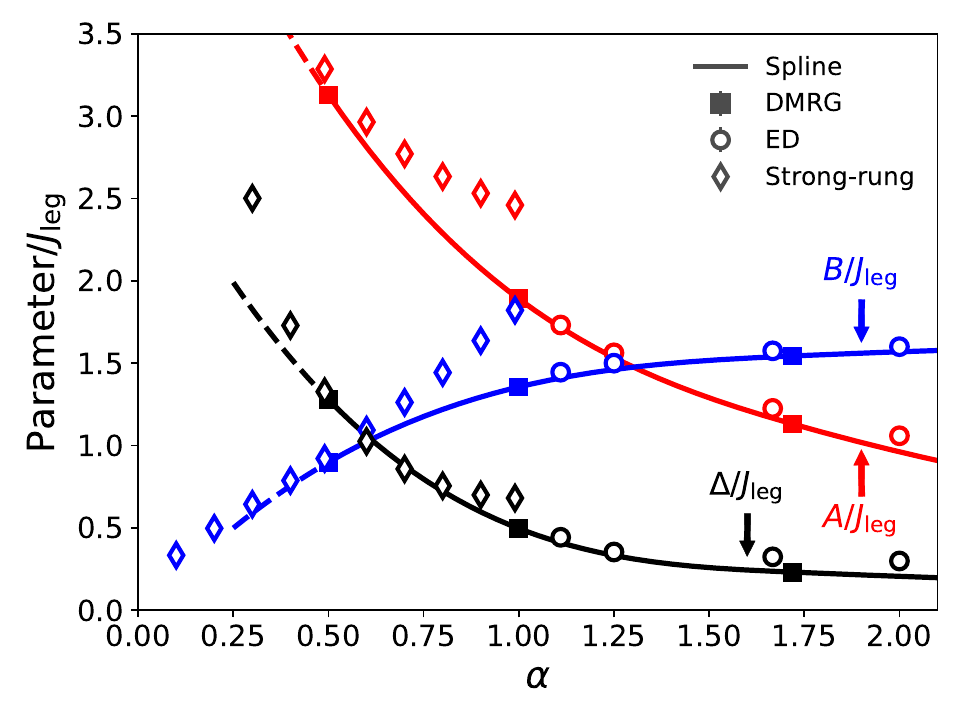}
    \caption{{\bf Parametrization of the triplon dispersion in the single-mode approximation (SMA).} Dependence on $\alpha$ of the three SMA parameters $\Delta$ (black), $A$ (red), and $B$ (blue). We parametrized the SMA as a function of $\alpha$ by a cubic spline interpolation based on the DMRG calculations of Ref.~\cite{Schmidiger2013}, which are shown as solid squares. The SMA parameters derived from our ED calculations are shown as the open circles and the parameters obtained within the strong-rung approximation as open diamonds.}
    \label{fig:app:parametrized_SMA}
\end{figure}

\begin{figure}[t]
    \centering
    \includegraphics[width=0.48\textwidth]{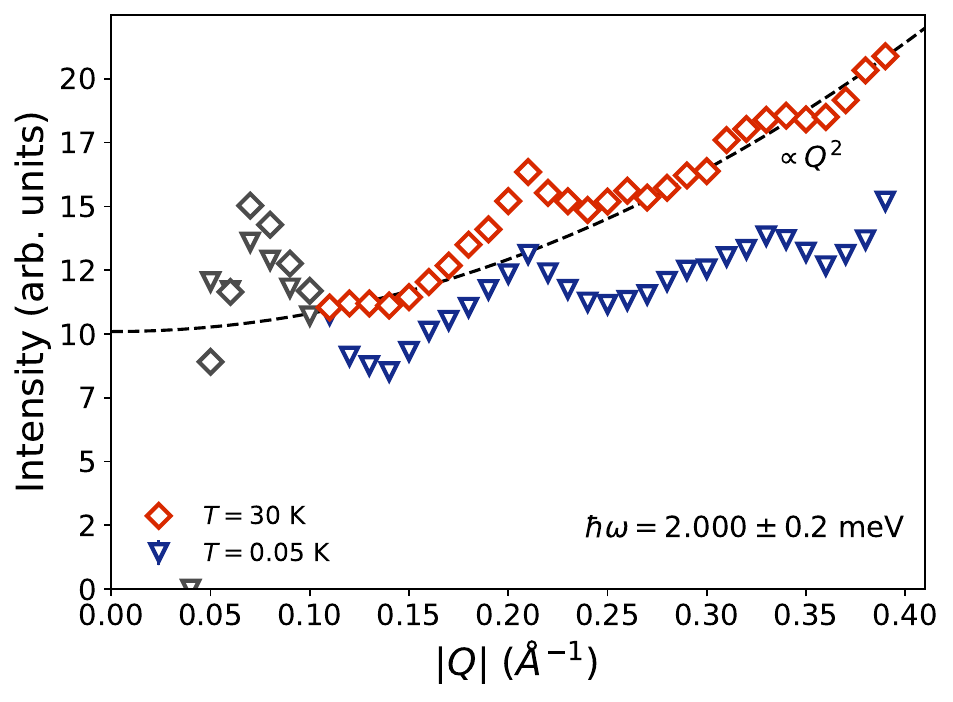}
    \caption{{\bf Phononic character of local modes.} Intensity of the vibrational modes obtained by integration over their full energy range and shown at two temperatures as a function of $|Q|$ to reveal conventional phononic behavior, excluding the possibility of a residual magnetic signal at these energies and temperatures. Gray symbols mark data excluded from the $|Q|^2$ fit due to the proximity of detector elements at these small $Q$ values to the direct beam.}
    \label{fig:powder_average}
\end{figure}

\section{Parametrized Single-Mode Approximation \label{app:sec:parametrized_SMA}}

The single-mode approximation (SMA) is a very robust, if heuristic, approach to fitting the primary features of a spectral function. Introduced in the context of spin ladders in Ref.~\cite{barnes1994}, it has since been used widely for the analysis of triplon dispersion relations \cite{Schmidiger2011}. The dispersion is expressed in the form 
\begin{equation}
\hbar\omega(k) \! = \! \sqrt{\Delta^2 \sin^2({\textstyle \frac12} k) \! + \! A^2 \cos^2({\textstyle \frac12} k) \! + \! B^2 \sin^2(k)},
\end{equation}
with the parameters $\Delta$ for the energy gap, $A$ for the energy at the zone boundary [$\hbar \omega(k = 0)$], and $B$ capturing the leading nontrivial contributions. To extract the $\alpha$ values for the two inequivalent spin ladders in Cu-CPA, we applied the SMA to the DMRG calculations shown in Fig.~6 of Ref.~\cite{Schmidiger2013} in order to obtain unbiased estimates of the three fit parameters $\Delta$, $A$, and $B$ over a range of $\alpha$ from 0.5 to 5. Based on these values, which are shown as the solid squares in Fig.~\ref{fig:app:parametrized_SMA} and reported for completeness in Table~\ref{tab:app:cu-cpa_spline_dmrg}, we implemented a cubic spline interpolation, shown as the solid lines in Fig.~\ref{fig:app:parametrized_SMA}, that allows us to reproduce the triplon dispersion over a continuous range of $\alpha$. 

\begin{table}[b]
    \centering
    \begin{tabular}{l | l l l}
    $\alpha$ & $\;$ $\Delta/J_\mathrm{leg}$$\;$& $\;$ $A/J_\mathrm{leg}$ $\;$& $\;$ $B/J_\mathrm{leg}$$\;$ \\
    \midrule
    0.50$\;\;$ &$\;$ 1.278(9) & $\;$ 3.128(8) & $\;$ 0.898(25) \\
    1.00$\;\;$ &$\;$ 0.495(8) & $\;$ 1.890(8) & $\;$ 1.355(9) \\
    1.72$\;\;$ &$\;$ 0.230(12) & $\;$ 1.130(8) & $\;$ 1.543(7) \\
    5.00$\;\;$ &$\;$ 0.068(21) & $\;$ 0.379(10) & $\;$ 1.619(5)\\
    \end{tabular}
    \caption{SMA parameters extracted from the DMRG calculations of Ref.~\cite{Schmidiger2013}. These parameters are shown as the squares in Fig.~\ref{fig:app:parametrized_SMA} and were used to calculate the cubic spline interpolation shown as the solid lines in Fig.~\ref{fig:app:parametrized_SMA}).}
    \label{tab:app:cu-cpa_spline_dmrg}
\end{table}

To verify this parametrization, we performed ED calculations and fitted the dominant branch of the resulting spectra with the SMA to extracting a further set of $\Delta$, $A$, and $B$ parameters, which are shown as the open circles in Fig.~\ref{fig:app:parametrized_SMA}. We find good agreement with the cubic spline interpolations, with minor quantitative deviations reaching the 10\% level on $\Delta$ and $A$ (5\% on $B$) at larger $\alpha$, where the short ladder lengths treated by ED become more of a handicap. For this reason we relied on DMRG at larger $\alpha$, where multiple measurements on DIMPY ($\alpha = 1.72$) have established quantitative accuracy \cite{Schmidiger2011, Schmidiger2013M, schmidiger_emergent_2016}. At small $\alpha$, we compare our interpolation with results from the strong-rung regime by using Eq.~(8) of Ref.~\cite{Reigrotzki_strong-rung_1994} to obtain the $\Delta$, $A$, and $B$ parameters shown as open diamonds in Fig.~\ref{fig:app:parametrized_SMA}. As expected, the strong-rung approximation (applied in Ref.~\cite{Reigrotzki_strong-rung_1994} as an expansion to third order in $\alpha$) is quantitatively accurate in the range $\alpha < 0.5$, but shows appreciable deviations beyond this that result in qualitative changes to the shape of the triplon dispersion.

We mention here that the numerical calculations (DMRG and ED) treat the magnetic excitation in units of $J_\mathrm{leg}$. Thus in our parametrization procedure we do not fit $\hbar\omega$, but rather $\hbar\omega/J_\mathrm{leg}$, and hence work with the reduced parameters $\Delta/J_\mathrm{leg}$, $A/J_\mathrm{leg}$, and $B/J_\mathrm{leg}$ in Fig.~\ref{fig:app:parametrized_SMA}. Fitting our Cu-CPA data using the parametrized SMA therefore involves the additional step of reintroducing the energy scale, $J_\mathrm{leg}$, as a free parameter. By deducing the two ratios $\alpha_a$ and $\alpha_b$ from our experimental data and extracting the two energy units $J_{\mathrm{leg},a}$ and $J_{\mathrm{leg},b}$, we compute the two rung interactions $J_{\mathrm{rung},a}$ and $J_{\mathrm{rung},b}$ and hence obtain the Heisenberg terms reported in Sec.~\ref{sres}B as the complete description of the Cu-CPA system.

\begin{figure}[t]
    \centering
    \includegraphics[width=0.5\textwidth]{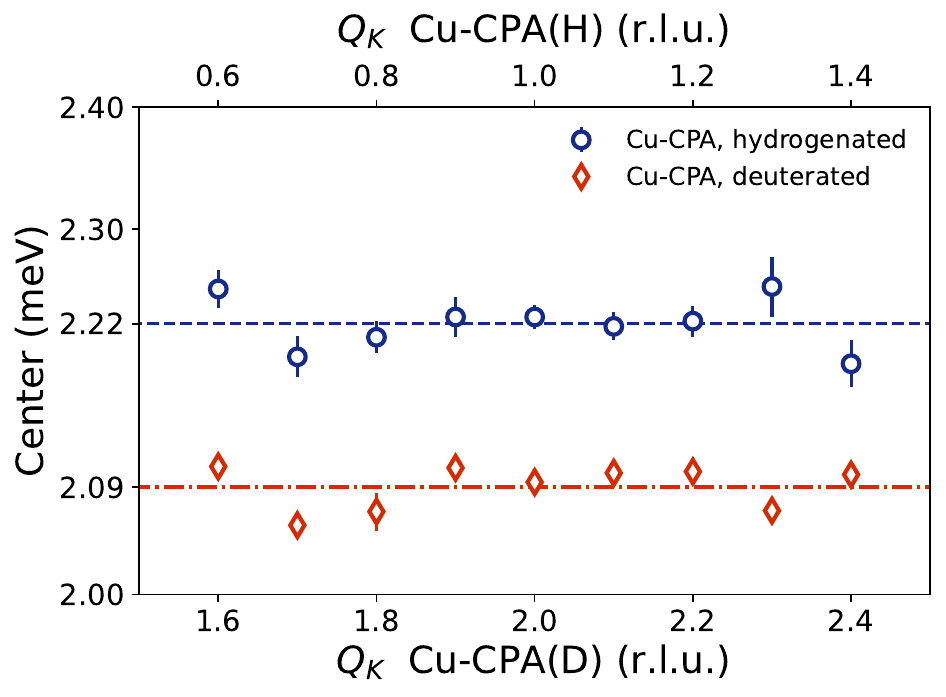}
    \caption{{\bf Isotope effect on the energy of the local phonon.} Fitted center position of the phonon mode appearing around 2.0~meV, shown as a function of $Q_K$ to verify its dispersionless nature. Blue circles show data for the hydrogenated Cu-CPA sample and red diamonds data for a deuterated one. The Cu-CPA(H) dataset was measured on LET and the Cu-CPA(D) dataset on IN5, both at 30~K. The horizontal lines mark the mean energies of the two phonons.} 
    \label{fig:CPA_center_flat_mode}
\end{figure}

\section{Properties of the local phonon}
\label{app:cpa_mass_vibration_mode}

To demonstrate that the flat modes appearing around 2~meV are of phononic origin, we prepared powder averages of the scattering intensity of our deuterated sample by integrating over the energy range 1.8--2.2~meV. As Fig.~\ref{fig:powder_average} shows, at 30~K this average exhibits clear $|Q|^2$ behavior, which is an unambiguous statement of its phononic nature.

To demonstrate that these phonons have their origin in the CPA molecules, we performed energy scans at $T = 30$~K over a range of $Q_K$-positions to compare the energies of the local phonon excitations in one hydrogenated and one deuterated sample of Cu-CPA. Because the phonons are non-dispersive, we integrated over all momenta in the perpendicular directions ($Q_H$ and $Q_L$). The center positions of each vibrational mode, determined by fitting a Lorentzian peak with a linear background, are shown in Fig.~\ref{fig:CPA_center_flat_mode}. Quite generally, the vibrational energy of a molecule is proportional to $1/\sqrt{m}$, where $m$ is the molecular mass. The ratio between the centers of the measured energies is 1.06(1), which is consistent with the square root of the mass ratios of both the partially and fully deuterated CPA molecules listed in Table~\ref{tab:CPA_masses_isotops}. Here we show the masses and mass ratios of CPA ligands without deuteration and with both partial and full deuteration, where partial deuteration refers to the replacement of 3/4 of the H atoms (the 9 H atoms bonded to C but not the 3 bonded to N), because this is more readily achieved than full deuteration.

\begin{table}[b]
    \centering
    \begin{tabular}{l c c c c}
        Atom & Occurrence & Mass (g/mol) & Total mass & $m^*$ \\
        \midrule
        C & 5 & 12.011 & 60.055 &\\
        H & 9 & 1.007&  9.063 &\\
        N & 1 &14.007&  14.007 &\\
        H & 3 & 1.007&  3.021 &\\
        \emph{Total}& & & \emph{86.146} & 1.000\\
        & & & &\\
        C & 5 & 12.011 & 60.055 &\\
        D & 9 & 2&  18 &\\
        N & 1 &14.007&  14.007 &\\
        H & 3 & 1.007&  3.021 &\\
        \emph{Total}& & & \emph{95.083} & 1.051\\
        & & &  &\\
        C & 5 & 12.011 & 60.055 &\\
        D & 9 & 2&  18 &\\
        N & 1 &14.007&  14.007 &\\
        D & 3 & 2&  6 &\\
        \emph{Total}& & & \emph{98.062} & 1.067\\
    \end{tabular}
    \caption{Mass summary for non-deuterated, partially deuterated, and fully deuterated CPA molecules, with the mass ratio $m^* = \sqrt{m_\mathrm{CPA}^\mathrm{(X)}/m_\mathrm{CPA}^\mathrm{(H)}}$ shown in the right-hand column.}
    \label{tab:CPA_masses_isotops}
\end{table}

\begin{figure}[t]
    \centering
    \includegraphics[width=0.5\textwidth]{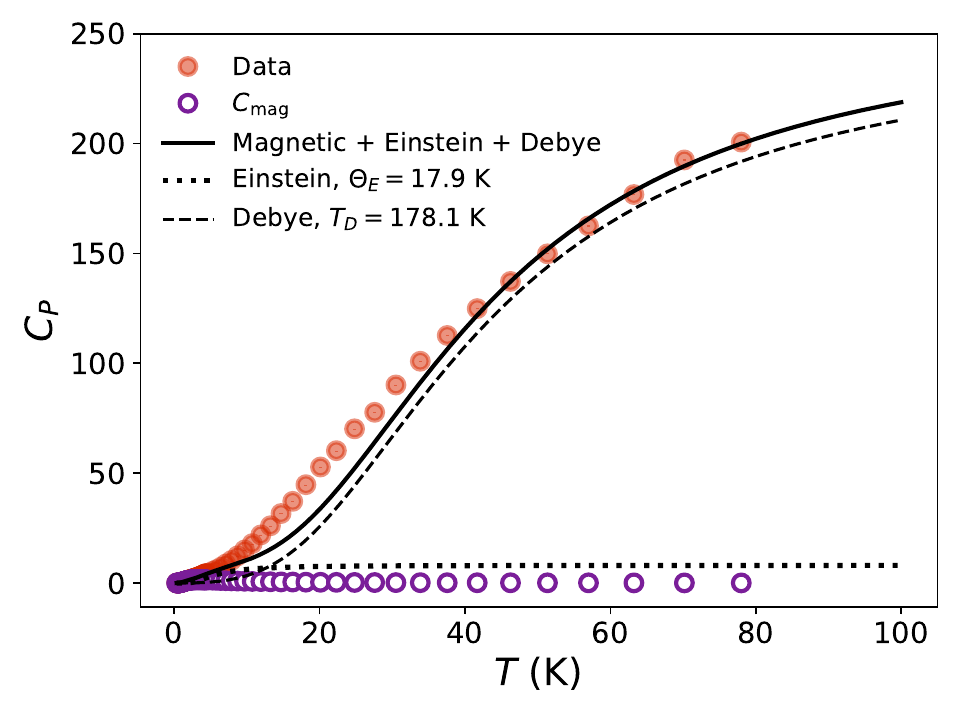}
    \caption{Specific-heat measurements of Ref.~\cite{philippe_metal-organic_2024} (red circles) fitted with a model including the magnetic contribution (purple circles), a Debye phononic contribution that accounts for the high-temperature part (dashed line), and an Einstein phononic contribution that accounts for the low-energy local phonon (dotted line). The total of these three contributions is shown by the solid line.}
    \label{fig:cu_cpa_specific_heat_full_range}
\end{figure}

\section{High-temperature specific heat}
\label{app:sh}

In Fig.~\ref{fig:cu_cpa_specific_heat_full_range} we show our fit to the specific-heat data reported in Ref.~\cite{philippe_metal-organic_2024} over the full range of available temperatures. We reiterate that full fit includes the magnetic specific heat computed from the ladder parameters determined by INS, the contribution of a low-energy local phonon modelled as an Einstein mode (harmonic oscillator), and a generic phonon contribution modelled by fitting a Debye form to the high-temperature regime. The parameters of the two phononic components were fitted recursively. We note the deviation of the full fit from the data between 10 and 30~K, which suggests the influence of further local phonon modes of the CPA rings, but avoid the underconstrained exercise of fitting these. 
 
\bibliography{CPAbib}

\end{document}